\documentclass[preprint,3p]{elsarticle}

\usepackage{amssymb}
\usepackage[utf8]{inputenc}

\usepackage[export]{adjustbox}
\usepackage{booktabs}
\usepackage{amsfonts}
\usepackage{graphicx}
\usepackage{amsmath, amssymb}
\usepackage{listings}
\usepackage{bm}
\usepackage{floatrow}
\usepackage{hyperref}
\usepackage{booktabs}
\usepackage{subfig}
\usepackage{wasysym}

\newfloatcommand{capbtabbox}{table}[][\FBwidth]

\newcommand{\mean}[1]{\langle #1 \rangle}

\newcommand{\eps}{\epsilon}
\newcommand{\rey}{\mbox{Re}}
\newcommand{\ttt}[1]{\texttt{#1}}

\begin{document}
\title{Predictive Accuracy of Wall-Modelled Large-Eddy Simulation on Unstructured Grids}

\author[focal1]{T.~Mukha\corref{cor1}}
\ead{timofey@chalmers.se}
\author[focal1]{R.E.~Bensow\corref{cor2}}
\ead{rickard.bensow@chalmers.se}
\author[focal2]{M.~Liefvendahl}
\ead{mattias.liefvendahl@foi.se}
\cortext[cor1]{Principal Corresponding Author}
\cortext[cor2]{Corresponding Author}
\address[focal1]{Chalmers University of Technology, Department of Mechanics and Maritime Sciences, 412 96 Gothenburg, Sweden}
\address[focal2]{FOI, Totalf\"{o}rsvarets forskningsinstitut, 164 90 Stockholm, Sweden}


\begin{abstract}
The predictive accuracy of wall-modelled LES is influenced by a combination of the subgrid model, the wall model, the numerical dissipation induced primarily by the convective numerical scheme, and also by the density and topology of the computational grid. The latter factor is of particular importance for industrial flow problems, where unstructured grids are typically employed due to the necessity to handle complex geometries. Here, a systematic simulation-based study is presented, investigating the effect of grid-cell type on the predictive accuracy of wall-modelled LES in the framework of a general-purpose finite-volume solver. Following standard practice for meshing near-wall regions, it is proposed to use prismatic cells. Three candidate shapes for the base of the prisms are considered: a triangle, a quadrilateral, and an arbitrary polygon. The cell-centre distance is proposed as a metric to determine the spatial resolution of grids with different cell types. The simulation campaign covers two test cases with attached boundary layers: fully-developed turbulent channel flow, and a zero-pressure-gradient flat-plate turbulent boundary layer. A grid construction strategy is employed, which adapts the grid metric to the outer length scale of the boundary layer. The results are compared with DNS data concerning mean wall shear stress and profiles of flow statistics. The principle outcome is that unstructured simulations may provide the same accuracy as simulations on structured orthogonal hexahedral grids. The choice of base shape of the near-wall cells has a significant impact on the computational cost, but in terms of accuracy appears to be a factor of secondary importance.

\end{abstract}

\maketitle

\section{Introduction} \label{sec:intro}

The majority of the computational fluid dynamics (CFD) solvers used for tackling flow problems occurring in applications are based on finite volume discretization of the governing equations.
As with other grid-based methods, the appropriate resolution and quality of the computational grid are crucial for obtaining an accurate solution.
In the simplest of cases, the mesh can be fully defined by a single parameter prescribing the cell size.
However, more often than not the domain of the simulation involves complex geometric features and discretizing it is a challenging and time-consuming task.
The cells of a typical mesh vary in size, shape, degree of anisotropy, etc.
Studies analysing and improving the accuracy of the finite volume method on such unstructured meshes can be found in the literature.
This includes works on truncation error analysis~\cite{Juretic2010, Kallinderis2015},
improving accuracy on particular types of grids~\cite{Nishikawa2020}, comparative evaluation of discretization scheme performance~\cite{Jalali2014}, and more practical studies looking at the grid effects on simulations of particular flows~\cite{Ghoreyshi2016}.


When the considered flow is turbulent, the appropriate grid construction strategy depends on the selected turbulence modelling approach.
Currently, the workhorse of industrial CFD is Reynolds-averaged Navier Stokes (RANS).
Here, when the flow is wall-bounded, it is considered best practice to cover the region occupied by the turbulent boundary layer (TBL) with prismatic layers.
This way the cells are aligned with the surface of the wall and the steep velocity gradient can be properly resolved.
In the rest of the domain the cell shape is not restricted.
Previously, the most common approach was to use triangular (tri-)prisms for the TBL and tetrahedra in the rest of the domain.
Recently, however, advances in meshing software made polygonal (poly-)prisms combined with arbitrary polyhedra an attractive option, due to the fact that this often results in a smaller number of higher-quality cells.
Finally, quadrilateral (quad)-dominant prismatic layers are also used, as well as hex-dominant meshes in the bulk flow.

Scale-resolving turbulence modelling approaches, such as direct numerical simulation (DNS) and large-eddy simulation (LES), have until recently mainly been used in academic studies of canonical flows.
The domains of such flows are well-suited for discretization with a high-quality structured grid.
This is reflected in existing guidelines for grid construction.
For example, based on analysis of integral length scales, it is typically recommended that in the TBL the spanwise resolution should be about twice as high as the streamwise.
Such a recommendation can be quite difficult to follow for a flow where the direction of the stream is subject to change, and for certain cell shapes may lead to deterioration of quality.
The DNS/LES community also appears to hold a degree of scepticism towards attempts at using these simulation techniques in conjunction with second-order numerical methods and unstructured grids.
As a result, the effects of the properties of unstructured grids on the predictive accuracy of DNS/LES have not been studied extensively.
For example, the only publication to directly compare the accuracy of DNS across several grid types appears to be the study of Komen et al.~\cite{Komen2014}.
Nevertheless, development of solvers for scale-resolving simulations on unstructured grids is an active area of research, see e.g.~\cite{Mahesh2004}.

It can be argued that with some exceptions both DNS and LES are still rarely used for anything but fundamental studies of turbulent flows, mainly due to the rapidly scaling costs of LES for flows with TBLs.
However, approaches that allow to alleviate this issue, such as hybrid LES/RANS and wall-modelled (WM)LES are used increasingly actively in applied research and industrial CFD.
Unfortunately, for these methods the dependence of the accuracy on the grid construction strategy is often even less predictable than for standard DNS/LES.
Establishing best practice meshing guidelines for these methods is therefore a pressing matter.

In the remainder of the paper we focus on grid construction strategies for WMLES.
For a detailed review of this turbulence modelling approach the reader is referred to~\cite{Larsson2016, Bose2018a}.
Several studies using WMLES with an unstructured grid can be found in the literature.
In most of them, the idea of covering the TBL region with prismatic cells is adopted.
In~\cite{Aljure2018}, WMLES is used to model a flow around a car model using tri-prism layers attached to the wall boundaries and tetrahedra in the bulk of the domain.
The same grid type is used in~\cite{Zhang2018a} to study the flow around a cylinder, and a rod-arifoil configuration in~\cite{Zhang2018}.
Flow around a displacement ship hull in model scale is considered in~\cite{Liefvendahl2018}.
Poly-prisms are used in the TBL region, and polyhedra elsewhere.
The NASA Juncture flow is investigated in~\cite{Lozano-Duran2019a}.
Here, a special grid construction approach referred to as `Voronoi hexagonal close packed point-seeding method` is used.
Based on the figures, the cells appear to be polyhedral.
Generally, the flow quantities analysed in these studies agree favourably with the reference data, indicating that it is possible to use an unstructured WMLES grid and achieve accurate results.

Initial efforts on carefully assessing how the properties of the unstructured grid affect the predictive accuracy of WMLES have been reported in~\cite{Mukha2018a}, which includes Mukha~and Liefvendahl~as co-authors.
The work presented here can be seen as a significant extension of this study.
The goal is to contribute towards finding the answers to the following questions.
What is a suitable unstructured grid construction strategy for WMLES?
What cell size metric is appropriate for comparing the resolution of grids consisting of cells of different shapes?
How much, if at all, does the accuracy of WMLES deteriorate when an unstructured grid is used instead of structured orthogonal hexahedra?
Excluding hexahedra, what cell shapes (or their combination) lead to the best accuracy?

Since these questions are very broad, here we limit the scope to attached TBLs, with fully-developed turbulent channel flow and a zero-pressure-gradient flat-plate TBL flow considered as test cases.
A grid construction strategy suitable for producing a high-quality mesh for WMLES of such flows is presented in Section~\ref{sec:tbl_meshing}.
Following the established practice, the boundary layer region is discretized with prismatic cells, with three candidate shapes for the base of the prisms considered: a quadrilateral, a triangle, and an arbitrary convex polygon.
The  corresponding grids are constructed to be equivalent with respect to an introduced grid resolution metric based on the average distance between the centre of a given cell and its neighbours, see Section~\ref{sec:grid_measures}.
The effects of having part of the TBL discretized with arbitrary non-primsatic cells are also evaluated.
For the case of channel flow, the effect of the resolution of the grid is also studied with 4 grid refinement levels considered.

To construct the grids we use Gmesh, Pointwise\textsuperscript{\textregistered} and Simcenter Star-CCM+\textsuperscript{\textregistered}.
The former is open-source, and the latter two are widely used in industry.
The open-source finite volume-based CFD software OpenFOAM\textsuperscript{\textregistered} 4.1 is used to run the simulations, coupled with an open-source library for wall-modelling~\cite{Mukha2019}.
To ensure the reproducibility of the study and also to broaden its impact, the article is supplemented by a rich dataset\footnote{DOI: \texttt{10.6084/m9.figshare.12482438}}, containing all the obtained results as well as ready-to-run OpenFOAM\textsuperscript{\textregistered} cases, including the associated meshes.

A brief summary of the contents of the remainder of the paper follows.
In Section~\ref{sec:grid}, a metric for the resolution of the WMLES grid is proposed, and the employed grid construction strategy is presented.
A short overview of surface meshing algorithms is given.
Section~\ref{sec:cfd} introduces the CFD methods that are used in the study.
The results of the conducted simulations are presented and discussed in Section~\ref{sec:results}, and concluding remarks are given in Section~\ref{sec:conclusions}.

\section{Unstructured grid generation} \label{sec:grid}
This section discusses how an unstructured grid suitable for WMLES of an attached TBL flow can be constructed.
In the first subsection, a metric for the grid resolution is introduced and related to properties of geometrical shapes that are commonly used in surface meshing.
Next, a strategy for constructing a boundary layer mesh is presented.
The final subsection is dedicated to the choice of the surface meshing algorithm.

\subsection{Grid resolution metrics} \label{sec:grid_measures}
A fundamental difficulty in comparing results from simulations using different grid cell shapes is defining a metric of grid resolution with respect to which the different grids are constructed.
In this work, the average distance between the centre of a given cell and the centres of its neighbours, $d$, is employed to that end.
Two grids with the same distribution of $d$ are thus considered equivalent in terms of resolution.
This approach is justified by the fact that $d$ directly enters the discrete formulation of the governing equations and, therefore, the measured grid resolution is explicitly connected to the discretization error.
Two alternative metrics that could be considered are the cell volume (or its cubic root) and the average length of the cell edges.
The former is often used for the LES filter width, whereas the latter is a common input parameter to grid generation algorithms.

It is instructive to compare these three metrics for three fundamental two-dimensional cell shapes: an equilateral triangle, a square, and a regular hexagon.
The relationships between the area $S$, the edge-size $a$, and $d$ (which for these elements coincides with the diameter of the inscribed circle) are given in Table~\ref{tab:fundamental_elements}.
It is evident that the choice of the resolution metric has a very significant effect on the comparative properties of the grids constructed using different elements.
For example, for a given $a$, covering a particular surface will require $\approx 6$ times more triangles than hexagons.
However, if $d$ is fixed, one would need approximately $1.5$ more polygons than triangles instead!

\begin{table}[htp!] 
	\begin{tabular}{lccc}
		\toprule
		& S(a)              & S(d)             & d(a)                         \\ \midrule
		Triangle & $\frac{\sqrt{3}}{4} a^2 \approx 0.43a^2$  & $3\frac{\sqrt{3}}{4}d^2 \approx 1.30d^2$             & $\frac{\sqrt{3}}{3}a \approx 0.57a$ \\
		Square   & $a^2$             & $d^2$            & $a$           \\
		Hexagon  & $3\frac{\sqrt{3}}{2} a^2 \approx 2.60 a^2$ & $\frac{\sqrt{3}}{2} d^2 \approx 0.87d^2$ & $\sqrt{3}a \approx 1.73a$	\\ \bottomrule
	\end{tabular}
	\caption{The relationship between the area $S$, the edge-size $a$, and the diameter of the inscribed circle $d$ for an equilateral triangle, a square, and a regular hexagon.}
	\label{tab:fundamental_elements}
\end{table}

In WMLES, the aim is to resolve the structures in the outer layer of the TBL, whereas the inner layer is accounted for by the wall model.
The resolution of the grid should thus be computed with respect to the local turbulent boundary layer thickness $\delta$.
When $d$ is used to characterize the cell size, the corresponding metric is thus $\delta/d$.
In the literature~\cite{Larsson2016, Liefvendahl2017b}, the resolution is often measured as the number of cells within a $\delta^3$-cube.
A similar metric based on $d$ is~$(\delta/d)^3$, but, as discussed above, the number of cells this corresponds to heavily depends on the shapes of the cells constituting the grid.

\subsection{Boundary layer meshing algorithm} \label{sec:tbl_meshing}

Using unstructured grids makes it possible to accurately control $d/\delta$ even when $\delta$ varies in space.
A reasonable choice is to keep it constant, so that the same level of turbulence resolution is maintained throughout the whole boundary layer.
A high-quality mesh for the TBL region can be constructed by extruding the surface mesh to form prismatic layers.
Based on these considerations, an unstructured grid generation algorithm suitable for WMLES has been introduced in~\cite{Mukha2018a}.
It consists of the following steps.
\begin{enumerate}
	\item Estimate the distribution of $\delta$ by running a RANS precursor or using approximative analytical relations, when such are available.
	\item Construct a surface mesh with the chosen distribution of $d/\delta$ on the wall boundary.
	\item Extrude prismatic layers between the wall and $\delta$, with the number of layers equal to $\delta/d$.
	\item Apply suitable discretization to the rest of the computational domain.
\end{enumerate}

Most of the grids used in this work are generated according to this procedure.
In some channel flow simulations, however, only part of the boundary layer is covered with prismatic cells with the goal of analysing the effect of such a change on the predictive accuracy of the simulation.
This is discussed in more detail in Section~\ref{sec:channel}.

\subsection{Selection of surface meshing algorithms} \label{sec:surface_meshing}

A crucial step in the mesh generation procedure defined above is generating a high-quality surface mesh that serves as basis for the extruded prismatic cells.
In this work, three types of cells are considered for constructing the surface mesh: triangles (tris), quadrilaterals (quads), and arbitrary polygons (polys).
For the latter two, indirect meshing methods~\cite{Remacle2013} are used, meaning that the mesh generation consists of two steps: constructing a suitable triangular grid and then converting it into a mesh consisting of either quads or polys, respectively.
In the case of a quad grid, recombination is applied to the triangles, whereas a polygonal grid is obtained as a dual of the triangulation.

Recombination refers to taking several grid cells and combining them into a single cell.
Here, two triangles are combined into a single quad.
This implies that a higher-quality grid is generated if the triangulation consists of right-angled triangles with catheti of equal length.
In the ideal case, two such triangles sharing the hypotenuse can be recombined into a square.
Then, the edge length of the quads, and consequently $d$, is equal to the length of the catheti in the underlying triangulation.
An excellent algorithm~\cite{Remacle2013} for generating triangulations suitable for recombination is implemented in the free mesh generation software Gmsh~\cite{Geuzaine2009}.
Furthermore, the recombination algorithm guarantees that the mesh consists only of quads~\cite{Remacle2012}.

\begin{figure}[htp]
	\centering
	\includegraphics[width=0.5\linewidth, center]{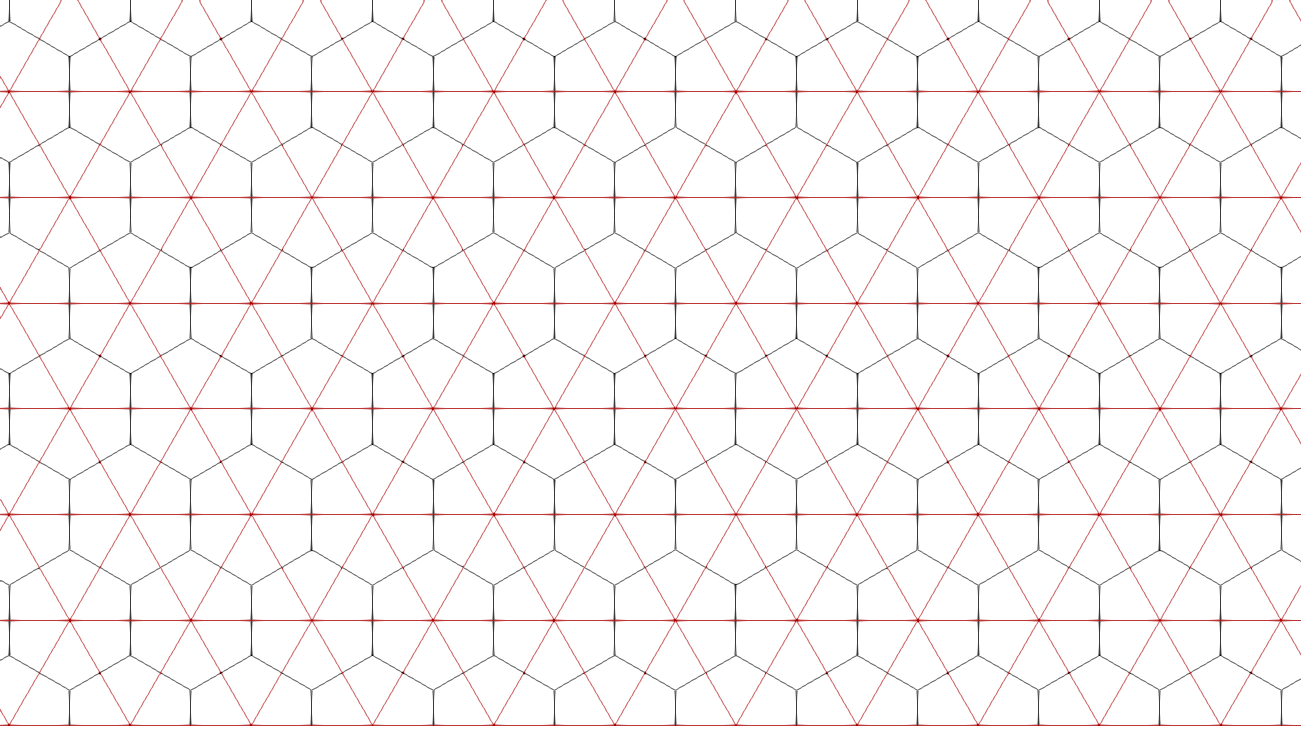}
	\caption{A surface triangulation and its dual. The domain boundary is shown at the bottom.}
	\label{fig:dual}
\end{figure}

A dual of a triangulation is a mesh in which the cell centres of the triangles are used as vertices, whereas the vertices of the triangles in turn become cell centres.
The edges of the dual mesh are constructed by connecting the cell centres of triangles that share an edge.
Special treatment is necessary at the boundary, where an edge may be built by connecting the cell centre of the triangle to the middle of the boundary edge.
An illustration of a triangulation and the corresponding dual is given in Figure~\ref{fig:dual}.
The figure also represents the case in which the triangulation is composed of equilateral triangles, which leads to a high-quality dual grid composed of regular hexagons.

\begin{figure}[htp]
	\centering
	\includegraphics[width=0.5\linewidth, center]{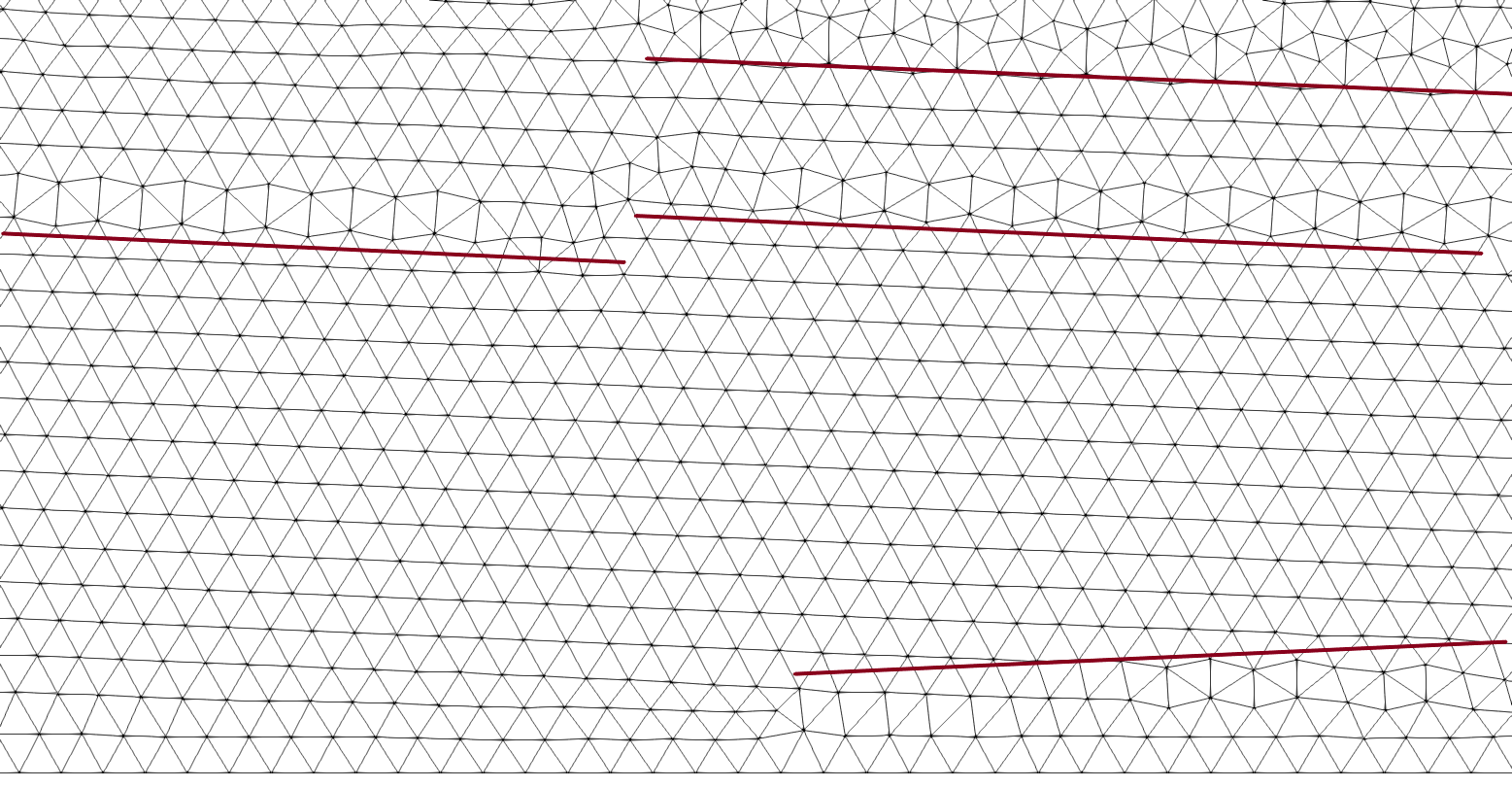}
	\includegraphics[width=0.5\linewidth, center]{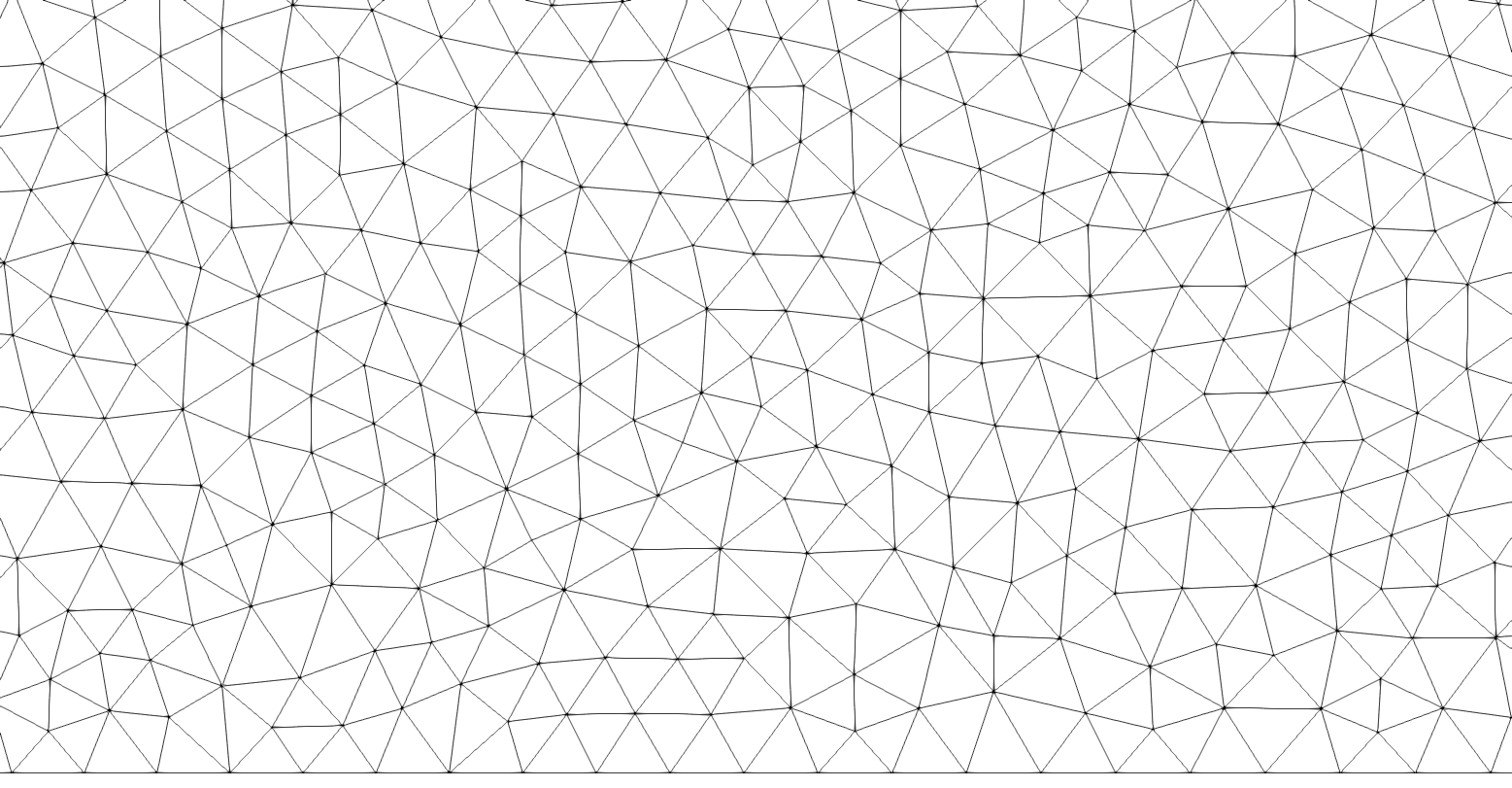}
	\caption{\textit{Top:} A triangulation generated with the Frontal Delaunay algorithm in Gmsh. Red lines highlight streaks of non-equilateral triangles. \textit{Bottom:} A triangulation generated with the Delaunay algorithm in Gmsh.}
	\label{fig:tri_for_poly}
\end{figure}

Several algorithms for generating triangulations suitable for conversion to polygonal meshes were tested.
In Gmsh, a Delaunay triangulation-based algorithm inspired by~\cite{Frey2008} is implemented.
The bottom plot in Figure~\ref{fig:tri_for_poly} shows a part of a flat surface mesh generated by this algorithm.
The quality of the grid is quite uniform throughout the surface, but it is easily seen that the resulting dual grid will contain polygons that are far from regular and having a varying number of edges.
Another algorithm that was considered is referred to as Frontal Delaunay in Gmsh, and based on the work of Rebay~\cite{Rebay1993}.
A grid generalized with Frontal Delaunay is shown in the top plot of Figure~\ref{fig:tri_for_poly}.
It mainly consists of equilateral triangles, however, elongated streaks of less regular triangles, highlighted with red lines in the figure, are also present.
Nevertheless, the quality of the triangles forming the streaks is acceptable, and, consequently, Frontal Delaunay is considered to be more suitable when generating a dual mesh.
Two algorithms included in the grid generation software Pointwise\textsuperscript{\textregistered} have also been tested.
The first one, referred to as Delaunay, produced meshes similar to the Delaunay algorithm in Gmsh.
The second one, called Advancing Front, generates the mesh sequentially starting from the boundaries.
More information on these type of algorithms can be found in~\cite{Schoberl1997} and the references therein.
Equilateral triangles dominate the generated grid, however, in places where two `fronts' propagating from separate boundaries meet, triangles of slightly lesser quality are present.
The result is thus quite similar to the grid produced by the Frontal Delaunay algorithm in Gmsh. 

Since both Fontal Delaunay and Advancing Front generate a surface mesh dominated by equilateral triangles, these algorithms are well-suited not only for constructing a dual poly-mesh, but also for the case when the triangles themselves are used as the bases of the prisms.

It is noted that besides for the quality of the mesh, the algorithms also differ in terms of execution speed and robustness.
In this work, planar wall surfaces are considered, discretized using a relatively low number of cells ($\lessapprox 5 \cdot 10^5$), which makes these factors unimportant.
However, in more practical cases they can become critical.
\section{Computational fluid dynamics methods} \label{sec:cfd}
In this work, the open source CFD software OpenFOAM\textsuperscript{\textregistered} version 4.1 is used to solve the LES equations.
OpenFOAM\textsuperscript{\textregistered} is based on cell-centred finite-volume discretization, and, crucial for this study, grids consisting of arbitrary convex polyhedral cells are supported.
The flow is considered incompressible and no explicit filtering is applied to the unknowns, i.e.~the velocity, and pressure fields.
Instead, the equations are considered to be implicitly filtered by the discretization procedure.
The WALE model~\cite{Nicoud1999a} is selected for modelling the subgrid scales based on its favourable performance in WMLES on structured meshes~\cite{Rezaeiravesh2019a, Temmerman2003}.

For wall-modelling, a separate publicly-available library is employed~\cite{Mukha2019}.
An algebraic wall model based on Spalding's law-of-the-wall~\cite{Spalding1961} is used.
This model is accurate for the flow cases considered here, since they are free of strong pressure gradients, and the TBLs always remain attached.
For each wall-face and at each time-step the wall model samples the value of velocity from the LES solution at some distance $h$ from the wall, and uses the law-of-the-wall to compute an estimate of the wall shear stress, $\tau_w$ (or equivalently the friction velocity, $u_\tau = \sqrt{\tau_w}$)\footnote{In the governing equations the pressure gradient and viscous stresses (including $\tau_w$) are normalized by the fluid density.
Following standard practice, we will nevertheless use the terms `pressure' and `stress' when referring to the corresponding mass-specific quantities in the remainder of the text.}.
The value of $h$ has a significant effect on the accuracy of the wall shear stress prediction~\cite{Kawai2012}, in particular, it is necessary to sample the velocity further out than the wall-adjacent cell.
Here, $h$ is always set to the distance to the centre of the second consecutive off-the-wall cell.

An important issue is the choice of the interpolation scheme for the convective flux.
Here, a weighted linear blending of linear interpolation and second-order upwind interpolation is used.
The weights of the two schemes are, respectively, $0.75$ and $0.25$.
Both of them are second-order accurate and unbounded, however the numerical diffusion introduced by the upwind scheme is enough to restrict the effect of any introduced non-physical oscillation.
This treatment of the convective terms has been shown to provide good results in previous WMLES studies on hexahedral meshes~\cite{Mukha2019, Rezaeiravesh2019a}.
Furthermore, it has proven to be stable enough for simulations on unstructured meshes presented in this work.

The other components of the employed numerical approach are as follows.
The diffusive fluxes are interpolated linearly, with an explicit correction applied to reduce errors caused by mesh non-orthogonality, see~\cite[p.~83]{Jasak1996}.
A second-order backward-differencing scheme is used for integration in time, see e.g.~\cite[p.~91]{Jasak1996} for the definition.
The PISO algorithm~\cite{Issa1986} is used for pressure-velocity coupling, with 3 coupling iterations per time-step.

In the simulations of a flat-plate TBL flow presented in Section~\ref{sec:tbl}, it is necessary to trigger the transition to turbulence.
Random volumetric forcing is used to that end.
The magnitude of the introduced source term, $S$, is set according to the following in each cell
\begin{align} \label{eq:trip}
	& S = \frac{U_{ref} I_{ref}}{\Delta t} \alpha_r.
\end{align}
Here,  $U_{ref}$ and $I_{ref}$ are reference values of velocity and turbulent intensity, $\Delta t$ is the time-step, $\alpha_r$ is a random number sampled from a uniform distribution covering the interval $[-0.5, 0.5]$.
The source term is to be applied similarly to a trip-wire in a physical experiment, i.e.~in a thin strip of cells close to the wall.

\section{Numerical experiments} \label{sec:results}

In this section, results from WMLES of two flows are presented: fully developed turbulent channel flow and flat-plate zero-pressure-gradient TBL flow.
Both flows are simulated using several types of unstructured grids, and in the case of channel flow different resolutions of the grids are also considered.
The focus of the analysis is kept on the comparison between the results using different grids rather than the accuracy of WMLES in general and the associated trends in the error profiles.
For a detailed discussion of the latter the reader is referred to~\cite{Rezaeiravesh2019a}.
The triple $u$, $v$, $w$ will be used to denote, respectively, the streamwise ($x$), wall-normal ($y$) and spanwise ($z$) components of (the implicitly-filtered) velocity.
Angular brackets are used to denote the average value in time and across statistically homogeneous spatial directions, and the prime symbol is used to denote the fluctuations with respect to the average.

\subsection{Fully developed turbulent channel flow} \label{sec:channel}
\subsubsection{Simulation set-up}
As the first test case, fully developed turbulent channel flow at $\rey_b = 125\, 000$ is considered.
Here, \mbox{$\rey_b = U_b\delta/\nu$} is the bulk Reynolds number of the flow, with  $U_b = 1$~m/s denoting the bulk velocity, $\delta = 1$~m the channel half-height, and $\nu = 8 \cdot 10^{-6}$~$\text{m}^2/\text{s}$  the kinematic viscosity.
The corresponding friction velocity-based Reynolds number, $\rey_\tau = \mean{u_\tau}\delta/\nu$, is $\approx 5200$ according to the DNS data~\cite{Lee2015}, which is used here as reference for computing errors in the obtained profiles of the flow quantities.
To set~$\rey_b$, a time-dependant source term is added to the streamwise momentum equation, which adjusts the velocity field to match the prescribed value of $U_b$.

The computational domain is of size $8\delta \times 2 \delta \times 6 \delta$.
The use of unstructured grids makes periodic boundary conditions more complicated to implement since there is no one-to-one mapping between the faces of the two patches coupled by periodicity.
In OpenFOAM\textsuperscript{\textregistered}, the so-called Arbitrary Mesh Interface (AMI) can be used to rectify this issue.
The AMI computes the face connectivity between the two patches and ensures that the numerical fluxes are mapped conservatively.

Solution fields for the same flow obtained in previous studies were mapped onto the constructed grids to serve as initial conditions.
To remove any associated artefacts, all simulations were first run for $37.5T_f$, where $T_f = 8\delta/U_b$ is the flow-through time.
Afterwards, the gathering of statistical data was started, and the simulations were allowed to run for another $125T_f$.
In all the simulations, the time-step $\Delta t$ was set to $6.25 \cdot 10^{-3}T_f$.
The temporal resolution of turbulent motions is thus grid-independent.
All the parameters of the simulations are summarized in Table~\ref{tab:channel_sim_params}.

\begin{table}[htp!]
	\begin{tabular}{@{}ll@{}}
		\toprule
		Bulk velocity, $U_b$                          & 1 m/s                                   \\
		Kinematic viscosity, $\nu$                    & $8 \cdot 10^{-6}$ $\text{m}^2/\text{s}$ \\
		Channel half-height, $\delta$                 & 1 m                                     \\
		Bulk Reynolds number, $\rey_b$                & $125\,000$                              \\
		Target friction Reynolds number, $\rey_\tau$  & $5\,200$            \\
		Domain size, $L_x \times  L_y \times L_z$     & $8\delta \times 2\delta \times 8\delta$ \\
		Flow-through time, $T_f$                      & 8 s                                     \\
		Time-step, $\Delta t$                         & $6.25 \cdot 10^{-3}T_f = 5 \cdot 10^{-3}$ s \\
		Averaging time                                & $125T_f$,  1000 s                       \\ \bottomrule
	\end{tabular}
	\caption{Parameters of the channel flow simulations.}
	\label{tab:channel_sim_params}
\end{table}

\subsubsection{Grid generation}
In terms of the grid generation procedure  defined in Section~\ref{sec:tbl_meshing}, channel flow can be seen as a limiting case of two TBLs that together occupy the whole domain and have a constant thickness $\delta$ equal to the channel half-height.
As consequence, the prismatic layers should occupy the whole domain.
Thus, given the surface mesh for the bottom wall, a three-dimensional grid can be obtained simply by building the appropriate number of layers between $0$ and $2\delta$ in the wall-normal direction.
The utility \ttt{extrudeMesh} included in OpenFOAM\textsuperscript{\textregistered} was used to that end.

\begin{figure}[hpt!]
	\centering
	\subfloat[Hexahedra]{\includegraphics[width=0.45\linewidth]{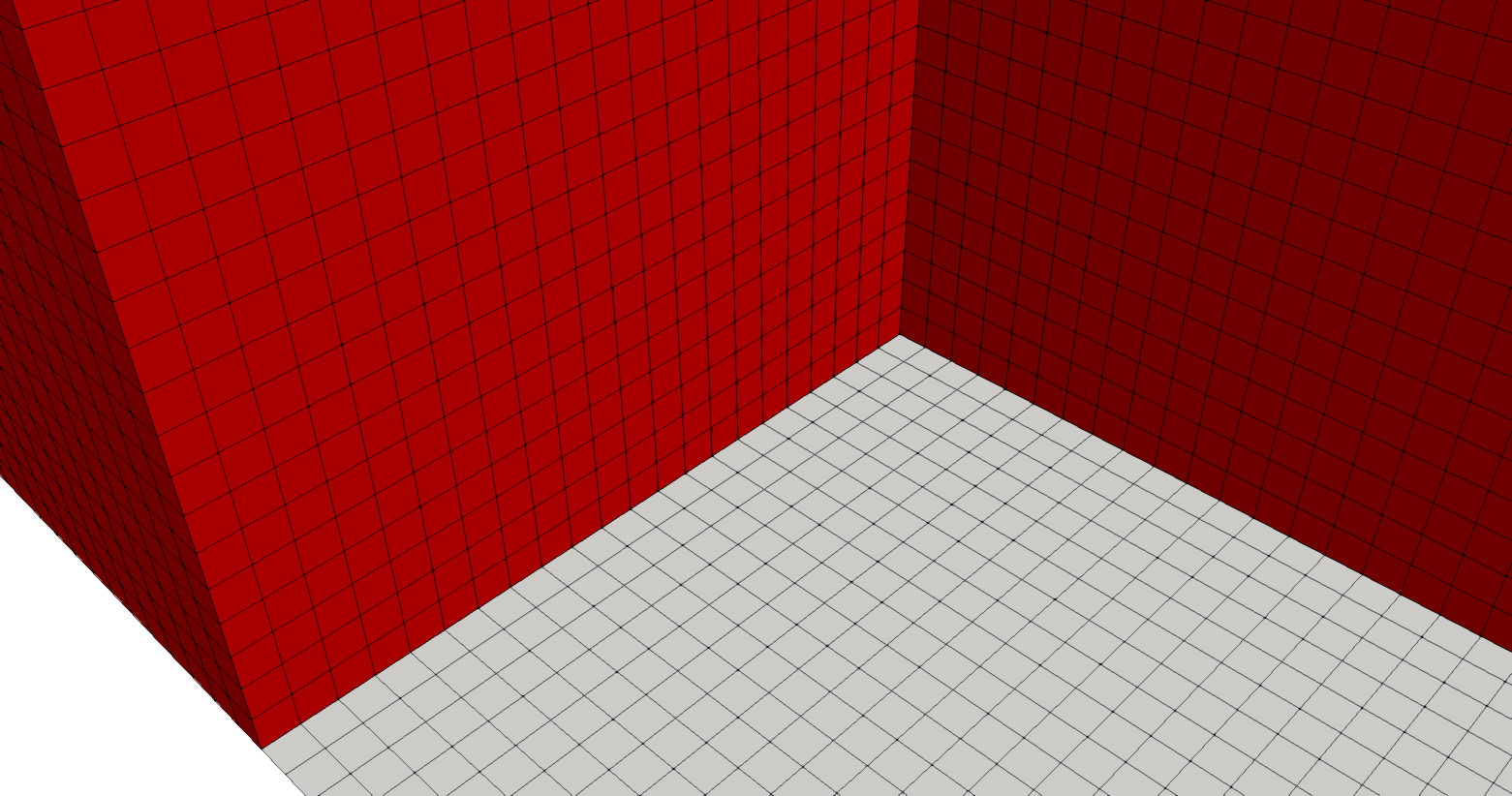}}
	\quad
	\subfloat[Poly-prisms]{\includegraphics[width=0.45\linewidth]{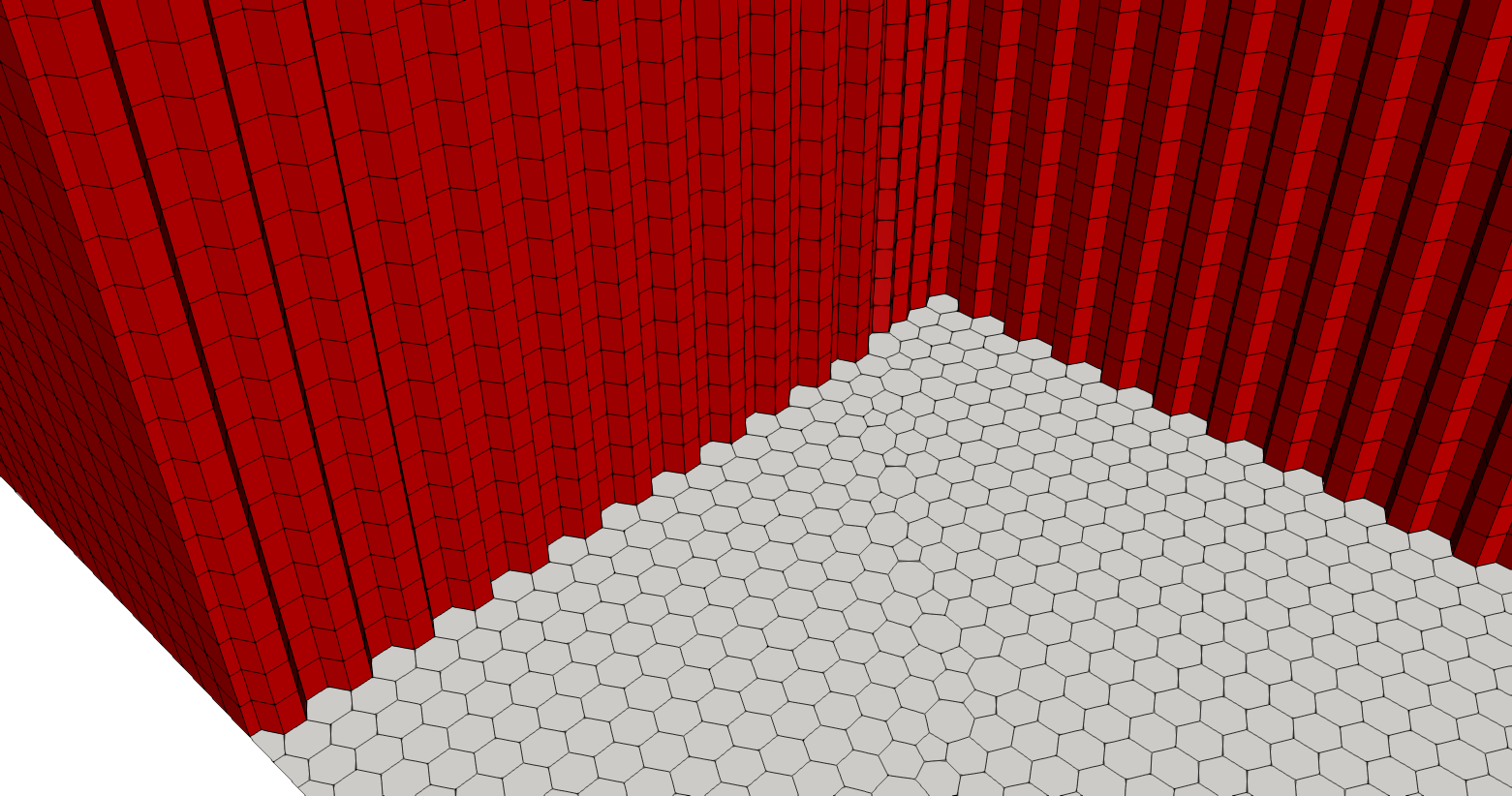}}
	\par
	\subfloat[Poly-hybrid]{\includegraphics[width=0.45\linewidth]{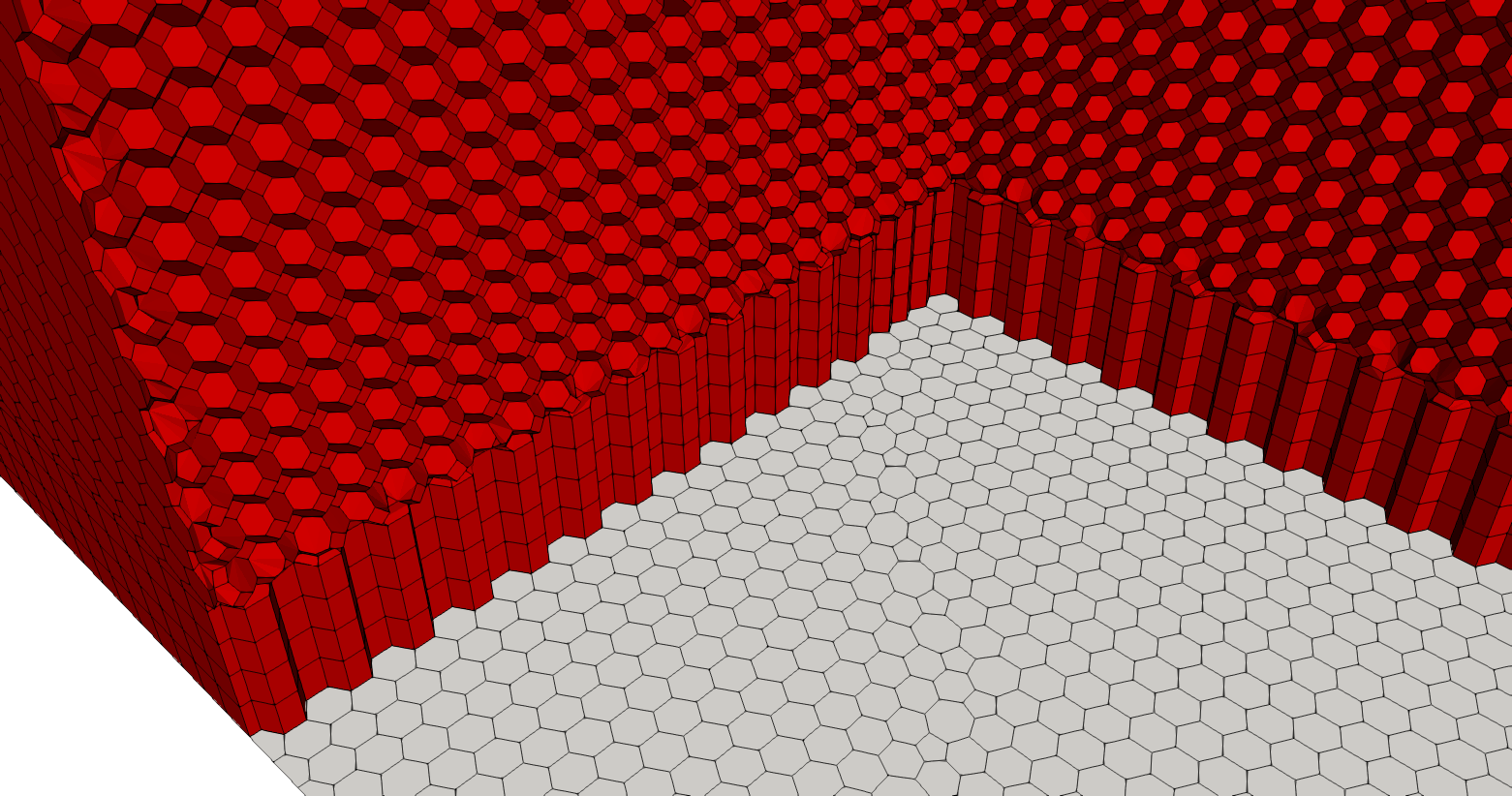}}
	\quad
	\subfloat[Tri-prisms]{\includegraphics[width=0.45\linewidth]{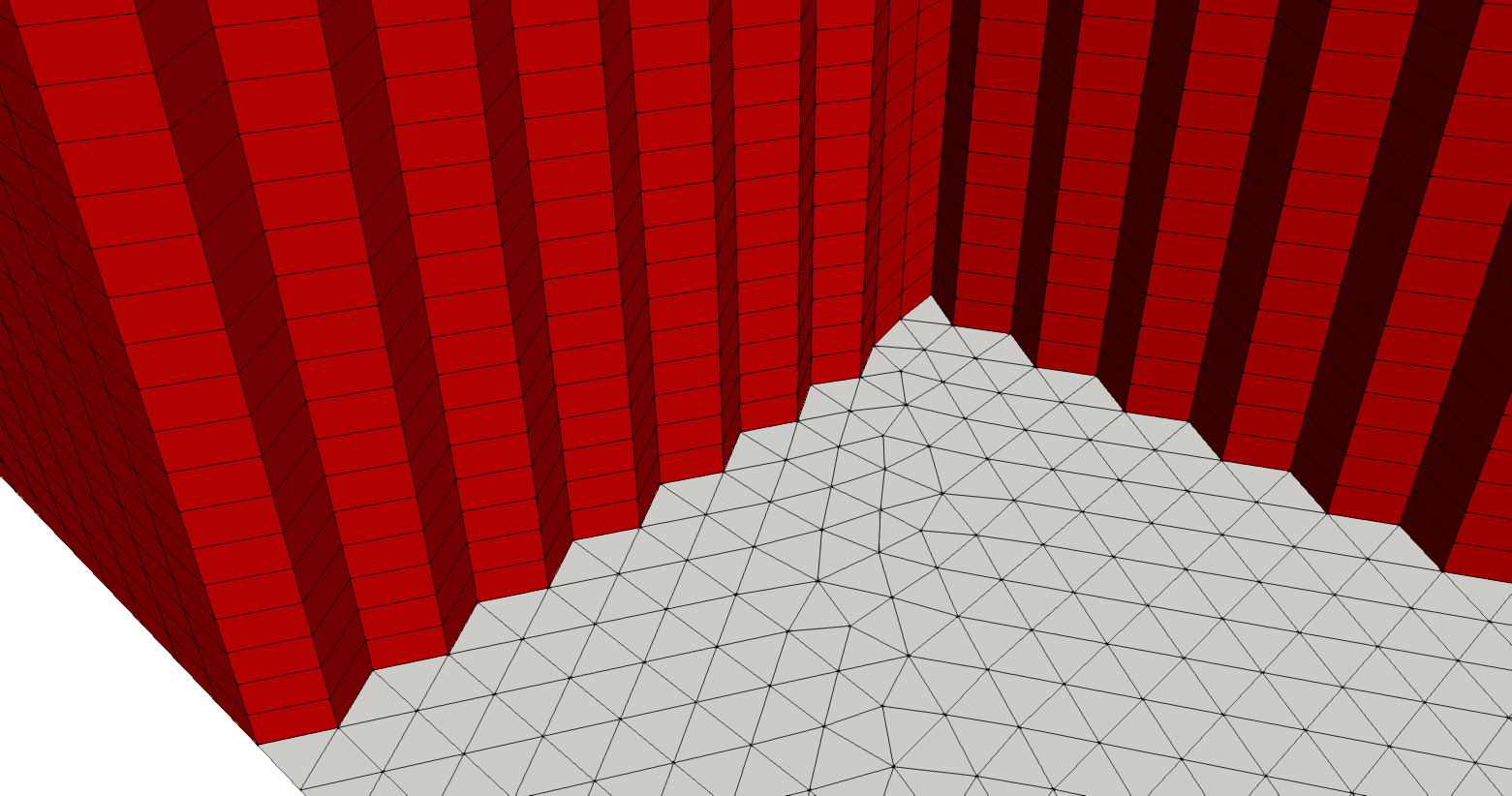}}
	\caption{Grid types used in channel flow simulations. In each plot, the grey elements indicate the wall-surface grid, and the volume grid is cut for visualization. The grids shown in the figure correspond to $d=1/20$.}
	\label{fig:channel_grids}
\end{figure}

As mentioned in Section~\ref{sec:surface_meshing}, tri-, quad- and poly-prism layers are considered in the study.
Note that in the case of channel flow, the quad-prism grid reduces to a structured hexahedral (hex) one.
The results from simulations on this mesh type are used to define a reference point for the accuracy that can be expected from the WMLES.
Additionally, Simcenter Star-CCM+\textsuperscript{\textregistered} was used to generate a grid in which poly-prisms extend only up to $0.2\delta$, with the rest of the domain meshed using arbitrary polyhedra.
This grid is referred to as poly-hybrid.
It is interesting to see if the simulation using the poly-hybrid grid will exhibit significant deterioration of accuracy, because in practical situations $\delta$ is known only approximately, and consequently a part of the TBL can be expected to end up outside the prismatic layers.
The same situation could also arise when the geometry is so complex that extruding 15-30 layers is impossible, but 3-5 layers could be constructed.
An illustration of the four different grid types considered is provided in Figure~\ref{fig:channel_grids}.

Besides for different cell shapes, different densities of the grids are considered: $d = \delta/15$, $\delta/20$, $\delta/25$, and $\delta/30$.
Thus, simulations were performed using a total of 16 different grids.
The number of cells, $N_\text{cells}$, in each grid is given in Table~\ref{tab:channel_grid_size}.
It is easy to verify that for the hex, tri- and poly-prism grids the sizes follow the analytical relationships presented in Section~\ref{sec:grid}, meaning that the triangles are close to equilateral, and the polys are predominantly regular hexagons.

A measure of the cost of the simulations, $N_\text{hours}$, is also reported in Table~\ref{tab:channel_grid_size}.
To obtain $N_\text{hours}$, log files were first used to compute the wall-clock time, in hours, it took to advance each simulation by one~$T_f$.
These numbers were than multiplied by the number of processors each respective simulation was run on.
Thus, $N_\text{hours}$ is an estimate of how many hours it would take to advance each simulation by one $T_f$ if the simulations were to be run in serial.
Note that the metric is exact only under the assumption of linear scaling of computational effort with the number of cores.
The scaling tests previously performed on the hardware used to run the simulations showed that within the range of core numbers employed in the study this assumption holds very well.
The data in Table~\ref{tab:channel_grid_size} shows clearly that the cell topology has a large effect on $N_\text{hours}$.
This is not only due to the difference in $N_\text{cells}$, but also due to the fact that the number of non-zeros in the system matrix scales with the average number of faces of each cell.
This is particularly vivid when looking at the values for the poly-hybrid grid.
For instance, on the coarsest level of resolution, the number of cells for this grid is comparable to that of the poly-prism one, but the cost is more than twice as large, $N_\text{hours} = 6.66$ versus 2.93.
The simulations on the tri-prism grid are, on the other hand, by far the cheapest.
For example, the simulation on the $\delta/30$ tri-prism mesh is less expensive than on the $\delta/20$ poly-prism mesh.



\begin{table}[]
	\begin{tabular}{lllllllll}
		\toprule
		$d$ & \multicolumn{2}{l}{Hex} & \multicolumn{2}{l}{Poly-prisms} & \multicolumn{2}{l}{Poly-hybrid} & \multicolumn{2}{l}{Tri-prisms} \\
		& $N_\text{cells}$  & $N_\text{hours}$ & $N_\text{cells}$      & $N_\text{hours}$  & $N_\text{cells}$  & $N_\text{hours}$ & $N_\text{cells}$     & $N_\text{hours}$     \\
		\midrule
		$\delta$/15       &  $0.365$ & 2.0  & $0.424$  & 2.9  & $0.554$  & 6.7  & $0.287$ & 1.2 \\
		$\delta$/20       &  $0.864$ & 5.2  & $1.000$  & 13.5 & $1.283$  & 18.1 & $0.652$ & 3.1 \\
		$\delta$/25       &  $1.688$ & 12.4 & $1.940$  & 16.2 & $2.781$  & 35.5 & $1.282$ & 7.3 \\
		$\delta$/30       &  $2.916$ & 18.6 & $3.356$  & 20.2 & $4.489$  & 63.0 & $2.218$ & 12.7 \\
		\bottomrule
		\caption{The number of cells, $N_\text{cells}$, in millions, in the grids used for channel flow simulations, and the number of hours, $N_\text{hours}$, it would take to advance each simulation by one $T_f$ on a single processor.}
		\label{tab:channel_grid_size}
\end{tabular}
\end{table}


\subsubsection{Results}
\begin{figure}[htp!]
	\centering
	\includegraphics[center]{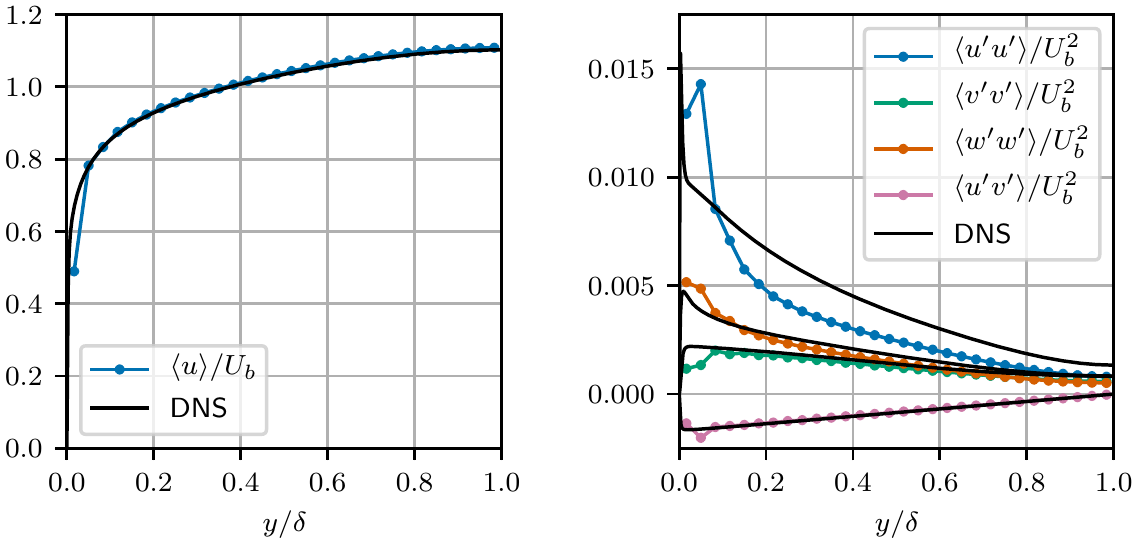}
	\caption{Results from the simulation on the $d = \delta/30$ hex mesh and the reference DNS data.}
	\label{fig:channel_overview}
\end{figure}

The results of the simulations are now presented and discussed.
For the majority of the quantities of interest, the obtained profiles will be presented in terms of the relative error, in percent, with respect to the reference DNS.
For example, for the mean streamwise velocity the profiles of
$$ \epsilon[\mean{u}/U_b] = \frac{[\mean{u(y)}/U_b]_{WMLES} - [\mean{u(y)}/U_b]_{DNS}}{[\mean{u(y)}/U_b]_{DNS}} \cdot 100$$
will be shown.
Since the size of WMLES cells are quite large at low $\delta/d$, the reference value at each cell centre is obtained by averaging the DNS profile across the wall-normal extent of the corresponding cell. 
The primary reason for not showing the profiles of the quantities themselves instead is that data corresponding to different simulations would be very difficult to distinguish in the plots.
However, to equip the reader with better intuition regarding the obtained level of predictive accuracy, scaled profiles of the first- and second-order velocity statistics are shown in Figure~\ref{fig:channel_overview} for the simulation conducted on the densest hex grid.

\begin{figure}[htp!]
	\centering
	\includegraphics[center]{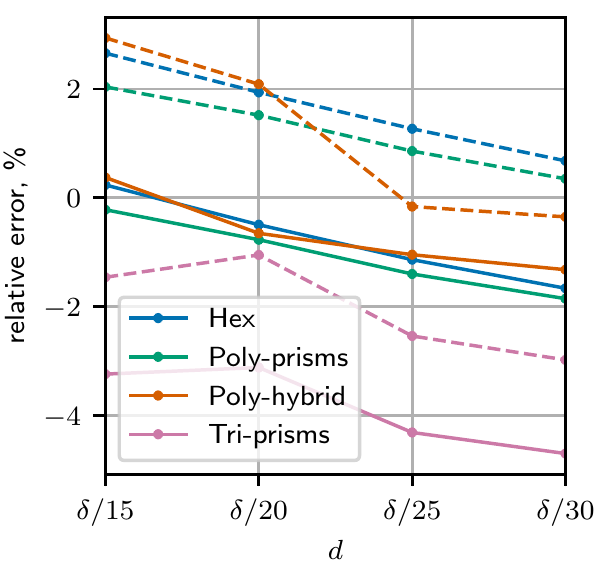}
	\caption{\textit{Solid lines:} Relative error in $\rey_\tau$. \textit{Dashed lines:} relative error in the mean streamwise velocity at the sampling point.}
	\label{fig:channel_utau}
\end{figure}

The performance of the wall model is now evaluated by considering the relative error in the obtained values of $\rey_\tau$, see Table~\ref{tab:channel_u_tau}.
Comparing the values corresponding to different meshes, the simulation on the tri-prism grid stands out as the least accurate, whereas using the other three grids results in a similar level of agreement with DNS, with the relative error not exceeding 2\%.
No monotonous improvement of the predictions of $\rey_\tau$ with the density of the grid is observed.
However, this is to be expected.
As extensively discussed in previous works~\cite{Mukha2019, Rezaeiravesh2019a}, for channel flow the error in the wall model's predictions is heavily correlated with the error in the velocity at the sampling point.
In Figure~\ref{fig:channel_utau}, both errors are plotted and indeed a clear connection between the two quantities can be observed.
Thus, for the error in $\rey_\tau$ to improve with mesh refinement, convergence in $\mean{u}$ at the sampling point is necessary.
Since in the current simulations the sampling point is fixed to the second off-the-wall cell, its location shifts linearly towards the wall as the mesh gets refined.
The characteristic size of the eddies in the log-law region, in turn, also scales linearly with the distance from the wall, meaning that their resolution at the sampling point remains effectively constant with mesh refinement.
For a more detailed discussion see~\cite{Kawai2012}.

\begin{table}[htp!]
	\begin{tabular}{ccccc}
		\toprule
		$d/\delta$  & Hex   & Poly-prisms & Poly-hybrid & Tri-prisms \\
		\midrule
		1/15        & 0.24   & -0.22       & 0.37      & -3.24      \\
		1/20        & -0.49  & -0.77       & -0.65     & -3.12      \\
		1/25        & -1.14  & -1.40       & -1.05     & -4.31      \\
		1/30        & -1.66  & -1.85       & -1.32     & -4.70      \\
		\bottomrule
	\end{tabular}
	\caption{Relative error in $\rey_\tau$ in simulations of turbulent channel flow.}
	\label{tab:channel_u_tau}
\end{table}

\begin{figure}[htp!]
	\centering
	\includegraphics[center]{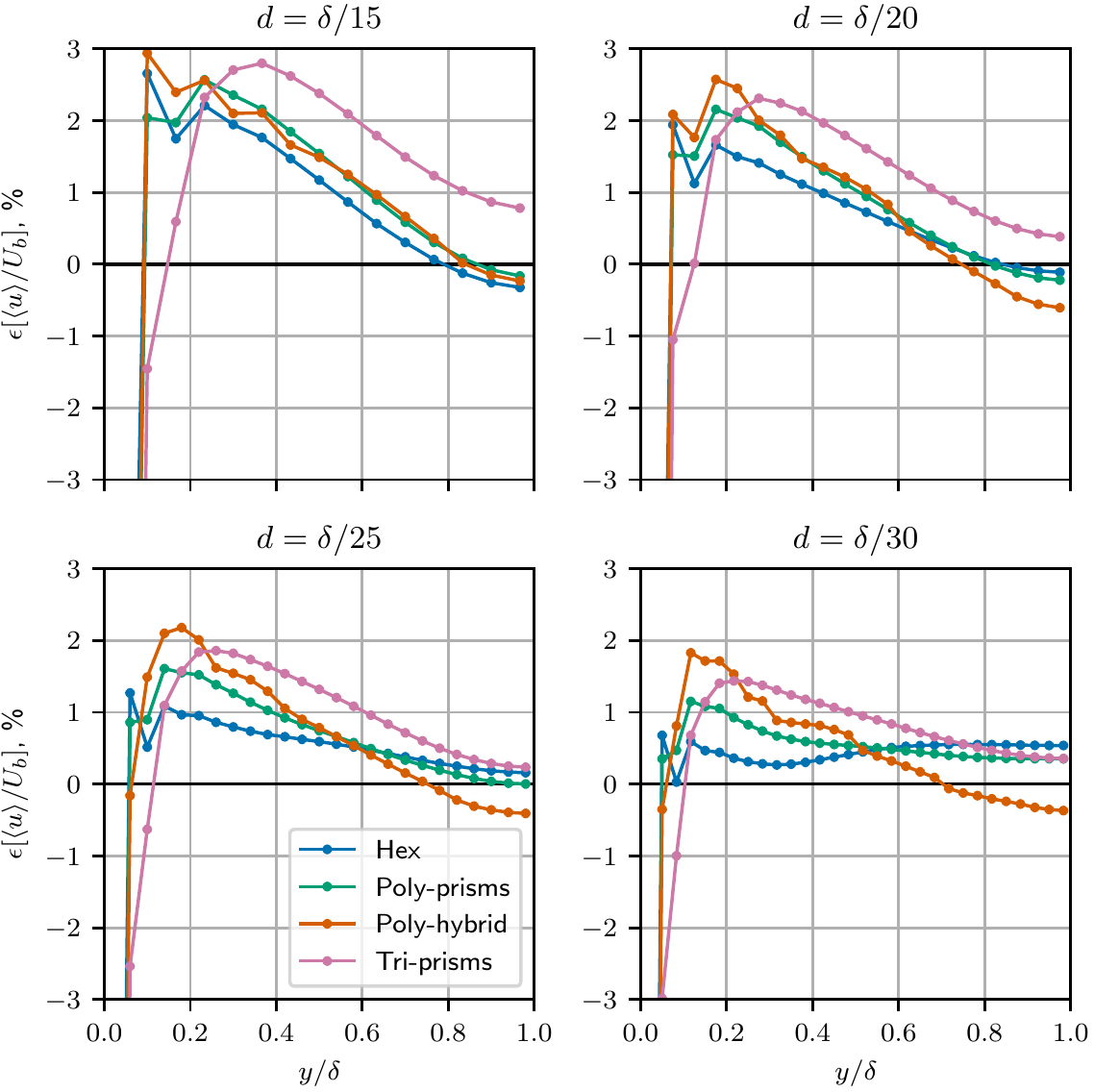}
	\caption{Relative error in the outer-scaled mean velocity profiles, $\eps[ \mean{u}/U_b]$, in simulations of turbulent channel flow.}
	\label{fig:channel_u}
\end{figure}

Figure~\ref{fig:channel_u} shows the relative error profiles of the outer-scaled mean velocity, $\eps[ \mean{u}/U_b]$.
Several trends previously reported in studies on hexahedral grids are observed.
The least accurate solution, heavily under-predicting DNS data, is always obtained in the wall-adjacent cell (the data point lies outside the ordinate range in the figure).
This is the fundamental reason behind not using this cell for sampling input into the wall model.
Excluding this point, the errors for all considered mesh resolutions and grids do not exceed 3\%.
For all four grids, some degree of convergence of the profiles with mesh refinement is present.
On the finest grids errors above the wall-adjacent cell do no exceed 2\%.
Overall, the solution on the hexahedral grid is more accurate than on the unstructured grids, but only slightly.
The least accurate profiles are obtained using the tri-prism grid, in particular for low values of $\delta/d$.

\begin{figure}[htp!]
	\centering
	\includegraphics[center]{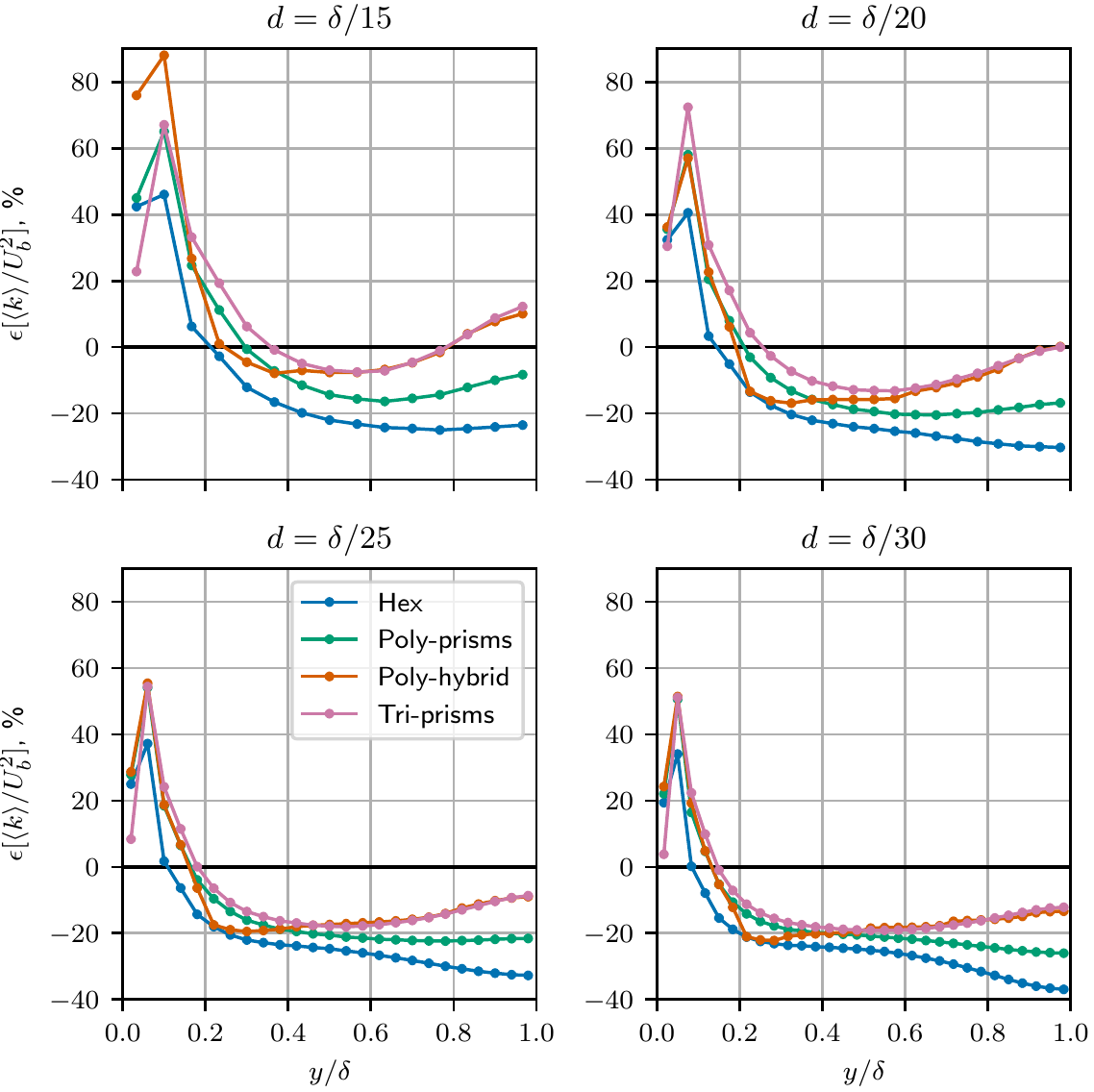}
	\caption{Relative error in the outer-scaled turbulent kinetic energy profiles, $\eps[ \mean{k}/U^2_b]$, in simulations of turbulent channel flow.}
	\label{fig:channel_k}
\end{figure}

Attention is now turned to the second-order moments of the velocity field.
First, the error profiles in the resolved turbulent kinetic energy is considered, see Figure~\ref{fig:channel_k}.
As expected based on previous works, the overall level of accuracy is significantly lower than what could be obtained for the mean velocity profile.
The largest errors occur in the inner region ($y \lessapprox 0.2$), where $k$ is heavily over-predicted.
Farther from the wall, under-prediction occurs instead.
Interestingly, the results on the structured grid are only better in the inner region, whereas in the core of the channel all the unstructured grids lead to a better result.
Presumably, this is due to some fortuitous cancellation of errors.

\begin{figure}[htp!]
	\centering
	\includegraphics[center]{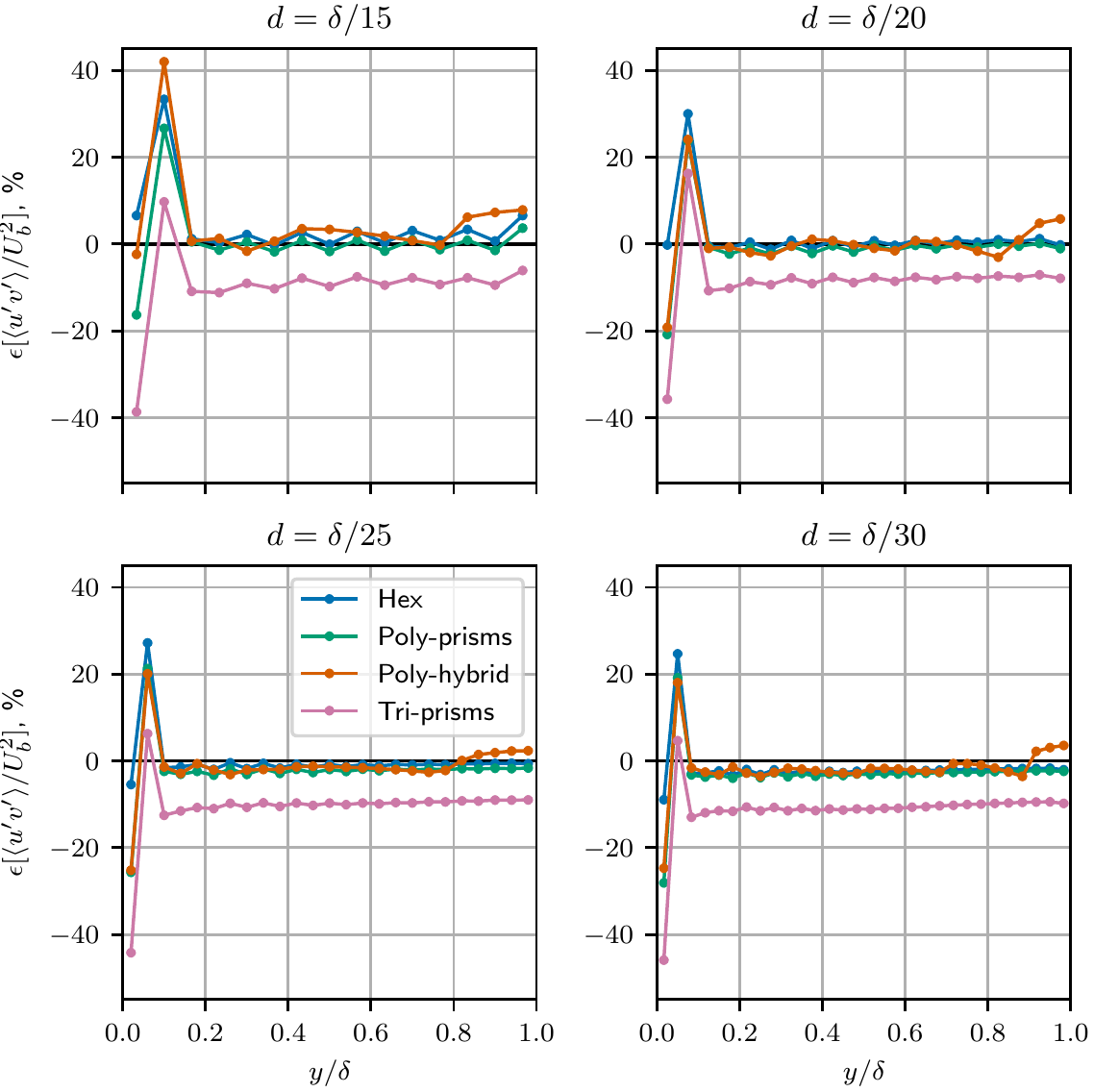}
	\caption{Relative error in the outer-scaled Reynolds shear stress, $\eps[ \mean{u'v'}/U^2_b]$, in simulations of turbulent channel flow.}
	\label{fig:channel_uv}
\end{figure}

The error profiles of turbulent shear stress are shown in Figure~\ref{fig:channel_uv}.
In the inner region, significant errors are observed independent of the type of mesh.
Above the logarithmic layer, where the wall-normal derivative of $\mean{u}$ is negligible, $\mean{u'v'}$ is not an independent quantity in the case of channel flow.
It is easy to show~\cite{Pope2000}, that $\mean{u'v'} \approx \mean{u_\tau}^2(y/\delta - 1)$.
The observed errors in this region are thus trivially related to the errors in mean friction velocity.
The somewhat strange behaviour of the error profiles close to the centreline of the channel is due to the fact that  $\mean{u'v'}(\delta) = 0$, and the relative error stops being a reliable measure.

Lastly, the power spectral density (PSD) of the velocity signal at $y/\delta \approx 0.5$ is considered.
To estimate the PSD, Welch's method was employed.
The sampled time-series of the velocity signal were divided into four segments with 50\% overlap, and the Hanning window applied to each segment.
A periodogram was then computed for each segment, and afterwards averaging across the four segments was applied.
Taking advantage of the spanwise periodicity of the solution, several time-series were acquired, corresponding to different spanwise locations.
To produce the final estimate of the PSD, an average across the estimates for each time-series was computed.
In the context of turbulent flows, a similar procedure was used in, for example,~\cite{Frohlich2005}.

\begin{figure}[htp!]
	\centering
	\includegraphics[center]{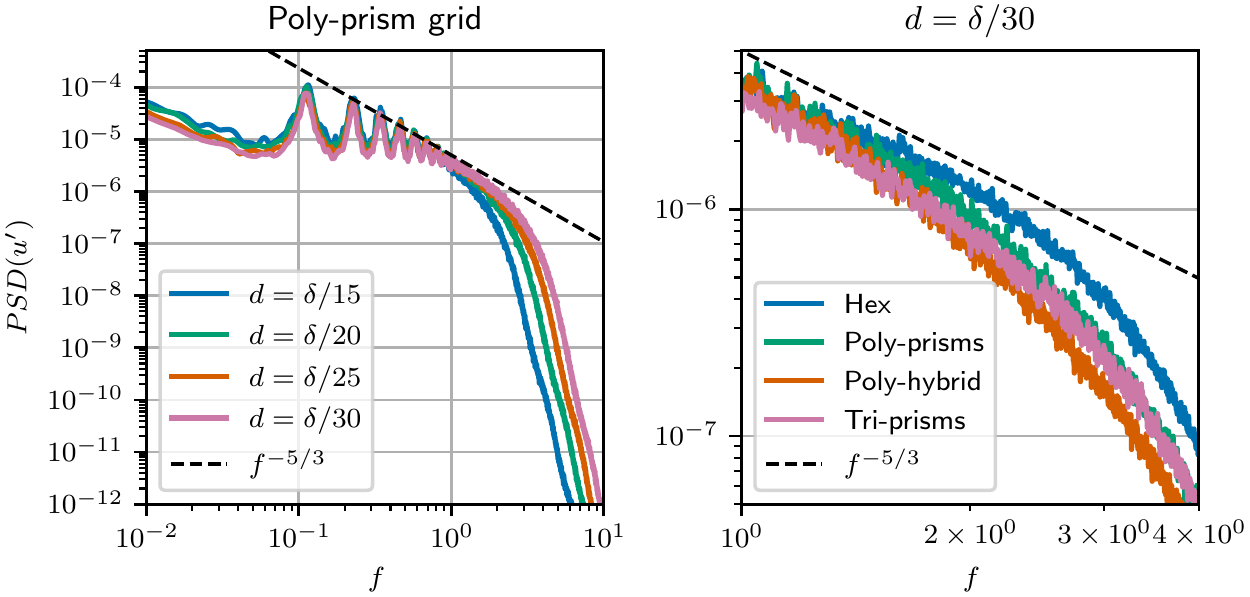}
	\caption{Power spectral densities of the streamwise velocity fluctuations.}
	\label{fig:channel_spectra}
\end{figure}

The PSDs of all three velocity components were computed for each of the conducted simulations, however, the results can be conveyed by looking at the streamwise velocity only, since the other two components exhibit very similar qualitative behaviour.
In the left plot of Figure~\ref{fig:channel_spectra}, the PSDs of $u'$ from the four simulations on the poly-prism grid are presented.
Virtually no difference in the results is observed in the frequency band corresponding to the most energetic eddies.
The pronounced peaks that can be seen in this region are due to the periodicity of the solution in the streamwise direction.
This is evident from the fact that the locations of the peaks are close to being multiples of $1/T_f$.
The effect of increased mesh resolution is visible in the $[1, 10]$ frequency band.
A portion of the band is occupied by the inertial range, as indicated by the $-5/3$-slope of the PSD.
At higher resolutions, the width of the inertial range increases due to the reduced amount of numerical dissipation.
However, even on the finest grid, it is very narrow compared to what can be expected at this Reynolds number.

In the right plot of Figure~\ref{fig:channel_spectra}, the PSDs of $u$ obtained in simulations on all the considered grid types are shown.
For all four, the finest grid is considered, $d = \delta/30$.
To be able to show the difference in the results, the plot focuses on a narrow frequency band containing the inertial range.
One can see that, as expected, using the hexahedral grid leads to the least amount of damping.
Among the unstructured grids the poly-hybrid grid is marginally more diffusive than the other two.
However, the overall difference in the exhibited level of numerical diffusion among the considered grids is clearly very small.

\subsubsection{Summary} \label{sec:channel_summary}

A concise summary of the findings of the channel flow study is now given.
The results on the four considered grid types are generally similar in their level of accuracy.
This allows to conclude that the employed grid density metric, $d$, is well suited for this type of study.
Further support for this is given by the PSD plots in Figure~\ref{fig:channel_spectra}, which show that the level of numerical diffusion is very similar on all the grid types.

Among the unstructured grids, the best results for a given $d/\delta$ were obtained using poly-prisms.
Recall that the poly-hybrid grid was considered to evaluate whether covering a part of the TBL with non-prismatic cells would lead to a significant deterioration of accuracy.
This turned out to not be the case, and the obtained accuracy is acceptable, even if subpar to that of the simulations on the poly-prism grid.
The results on the tri-prism mesh stand out in their relatively poor prediction of the mean friction velocity (and thus $\rey_\tau$), which, in turn, transitions into an even larger error in the turbulent shear stress.
Generally, the error in $\mean{u}$ is also somewhat larger when the tri-prism mesh is used.
But it should also be stressed that the simulations on the tri-prism mesh are significantly cheaper than on all the other meshes.
As already mentioned above, the $\delta/30$ simulation on the tri-prism mesh is cheaper than on the $\delta/20$ poly-prism mesh (see Table~\ref{tab:channel_grid_size}), and the $\mean{u}$ profile is better predicted in the former.
\subsection{Flat-plate turbulent boundary layer} \label{sec:tbl}
\subsubsection{Simulation set-up} \label{sec:tbl_setup}
The set-up of the TBL simulations is identical to that used in~\cite{Mukha2018a}.
A box of size \mbox{$L_x=2$ m}, \mbox{$L_y = 1$ m}, \mbox{$L_z= 0.2$ m} constitutes the computational domain.
At the inlet ($x = 0$), a uniform velocity profile $U_0 = 20.4$~m/s imposed, and transition is forced at $x = 0.1$ m by applying random volumetric forcing~\eqref{eq:trip}, as discussed in Section~\ref{sec:cfd}.
The reference velocity and turbulence intensity parameters in~\eqref{eq:trip} are set to $U_0$ and $0.025 \, \text{m}^2/\text{s}^2$, respectively.
The strip of cells where the source term is active is 2 cm long, 3 mm high and stretches across the whole domain in the spanwise direction.
For the purpose of mesh generation, the flow is considered fully turbulent at $x_0 = 0.4$ m.
This estimate is inserted into the following power law~\cite{Rezaeiravesh2016} to obtain the distribution of $\delta$ in the domain:
\begin{equation} \label{eq:power_law}
\rey_\delta = 0.1222 \rey_x^{0.8628}.
\end{equation}
Here $\rey_\delta = \delta U_0/\nu$ and $\rey_x = x U_0/\nu$, with $\nu = 1.65 \cdot 10^{-5}$ $\text{m}^2/\text{s}$.
For $x < x_0$ the thickness is considered constant for simplicity, \mbox{$\delta = \delta(x_0) \approx 8.1$} mm.

The maximum thickness of the TBL is predicted to be \mbox{$\delta_\text{max} \approx 3.2$ cm}, which leads to \mbox{$L_y/\delta_\text{max} \approx 31.25$} and \mbox{$L_z/\delta_\text{max} \approx 6.25$}.
These ratios are sufficiently high to avoid negative effects due to employed boundary conditions, which are periodic in the spanwise direction and symmetric at the top boundary.
At the outlet, a homogeneous Neumann condition is applied to velocity and the pressure is set to 0.

The time-step used in each simulation is $\Delta t = 2$ $\mu$s.
For the initial conditions, the velocity and pressure fields from the WMLES of the same flow reported in~\cite{Mukha2018a} are used.
Statistics were gathered over a period of $0.5$ s, corresponding to $\approx 5 T_{ft}$.
All the simulation parameters are summarised in Table~\ref{tab:tbl_params}.

\begin{table}[h!]
	\caption{Turbulent boundary layer simulation parameters.}
	\label{tab:tbl_params}
	\begin{center}
		\begin{tabular}{l  l}
			\toprule
			Estimated maximum TBL thickness, $\delta_\text{max}$ & $3.2$ cm \\
			Length of the domain, $L_x$ & 2 m $= 62.5\delta_\text{max}$ \\
			Height of the domain, $L_y$ & 1 m $ = 31.25\delta_\text{max}$ \\
			Width of the domain, $L_z$ & 0.2 m $= 6.25\delta_\text{max}$ \\
			Freestream velocity, $U_0$ & 20.4 m/s \\
			Kinematic viscosity, $\nu$ & $1.65 \cdot 10^{-5}$ $\text{m}^2/\text{s}$\\
			Flow-through time, $T_{ft}$ & $0.1$ s \\
			Time-averaging period & $0.5$ s $= 5T_{ft}$ \\
			Time-step, $\Delta t$ & $2 \cdot 10^{-6}$ s $= 2.04 \cdot 10^{-5}T_{ft}$ \\			\bottomrule
		\end{tabular}
	\end{center}
\end{table}

\subsubsection{Grid generation} \label{sec:tbl_grid}
The grid generation process follows the algorithm presented in~\ref{sec:tbl_meshing}, with quads, polys and tris considered for meshing the wall surface and arbitrary polyhedra used for meshing the freestream region.
Downstream of the numerical trip ($x > 0.1$ m), the resolution of the grid is set to $\delta/d \approx 15$, where $\delta$ is approximated according to the power law, as discussed above.
Unfortunately, higher resolutions could not be afforded due to limitations in the available computational resources.
Upstream of $x = 0.1$ m, the mesh is rapidly coarsened, since the flow in that region is laminar and two-dimensional.

\begin{figure}[htp!]
	\centering
	\includegraphics[]{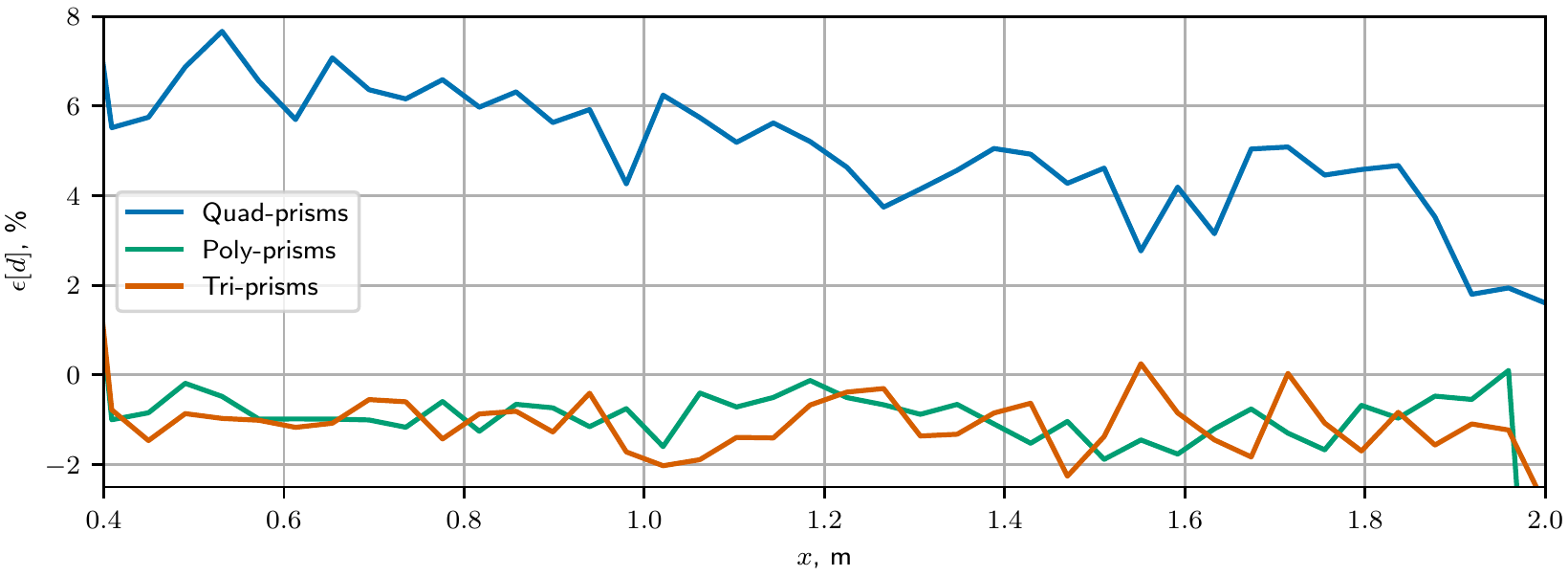}
	\caption{The relative error in $d$ with respect to the employed analytical estimate.}
	\label{fig:tbl_d}
\end{figure}

All three surface meshes were generated in Gmesh, using the algorithms discussed in Section~\ref{sec:grid}.
A particularly useful feature of Gmesh is the possibility to prescribe the distribution of the mesh size as an analytical functional relationship.
By that means, a formula based on~\eqref{eq:power_law} and the chosen value of $d/\delta$ could be used as input to the meshing algorithms, resulting in a very smooth variation of $d$ with $x$.
The relative error in the distributions of $d$ with respect to the analytical estimate is presented in Figure~\ref{fig:tbl_d}.
Unfortunately, for the quad-prism case there is, on average, a 5\% over-prediction.
This is most likely due to the fact that the input parameter to the surface meshing algorithm does not exactly correspond to the length of the catheti of the generated triangles.
However, since the power law~\eqref{eq:power_law} and the choice of $x_0$ are also approximate, this deviation is considered acceptable.
In fact, it will be shown below that the power law somewhat under-predicts $\delta$, meaning that for the quad-prism case the ratio of $d/\delta$ is reproduced slightly more faithfully.

\begin{figure}[htp!]
	\centering
	\subfloat[Quad-prisms]{\includegraphics[width=0.45\linewidth, trim={0 0 20cm 0}, clip]{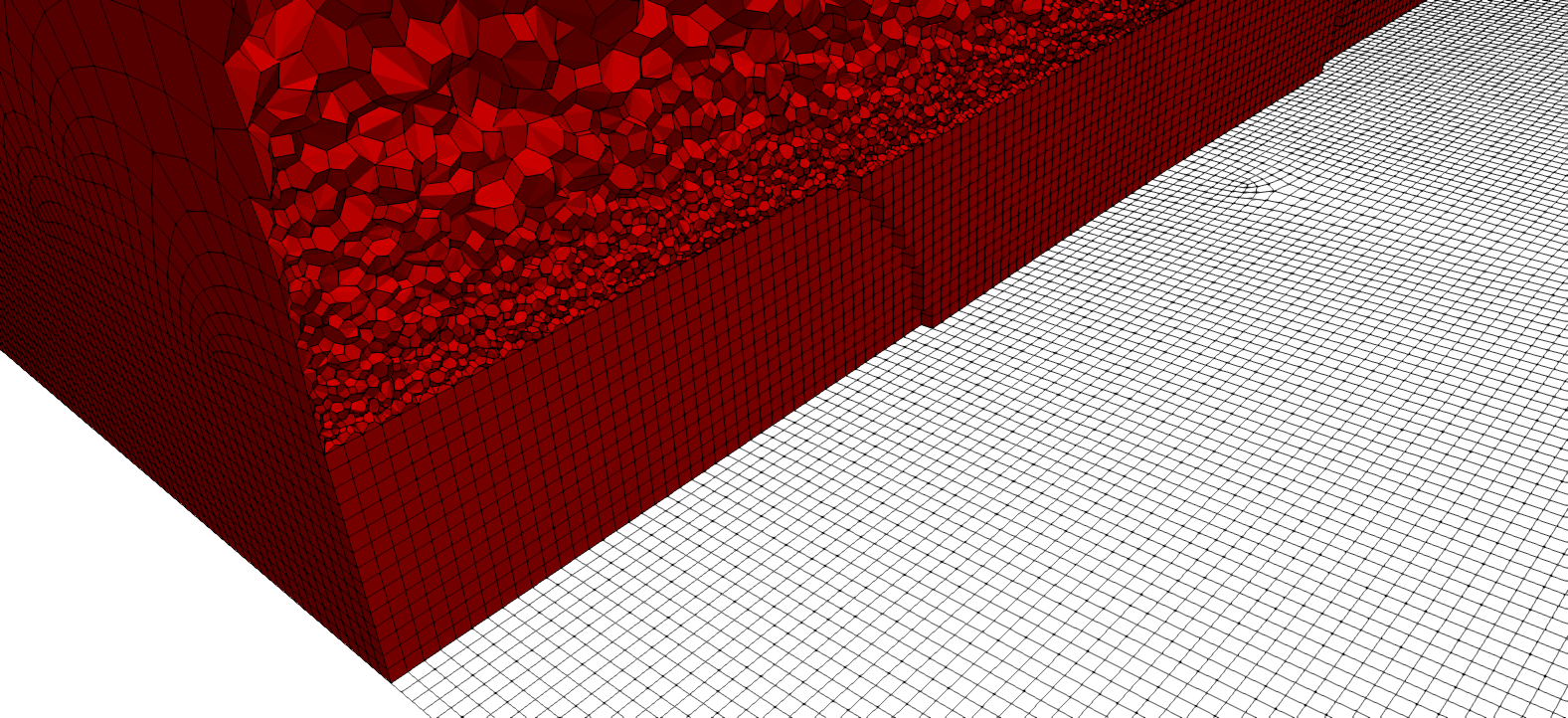}} \quad
	\subfloat[Poly-prisms]{\includegraphics[width=0.45\linewidth, trim={0 0 20cm 0}, clip]{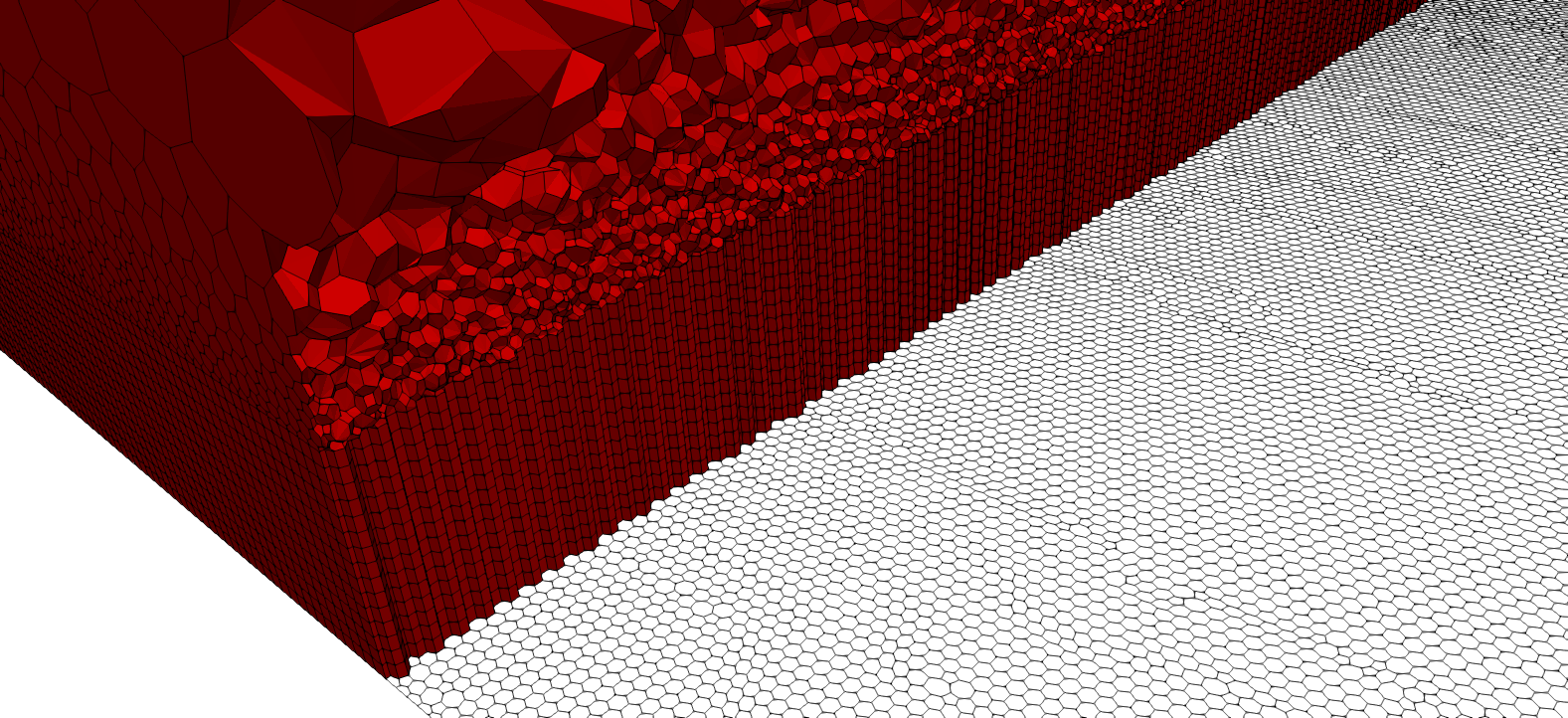}} \quad
	\subfloat[Tri-prisms]{\includegraphics[width=0.45\linewidth, trim={0 0 20cm 0}, clip]{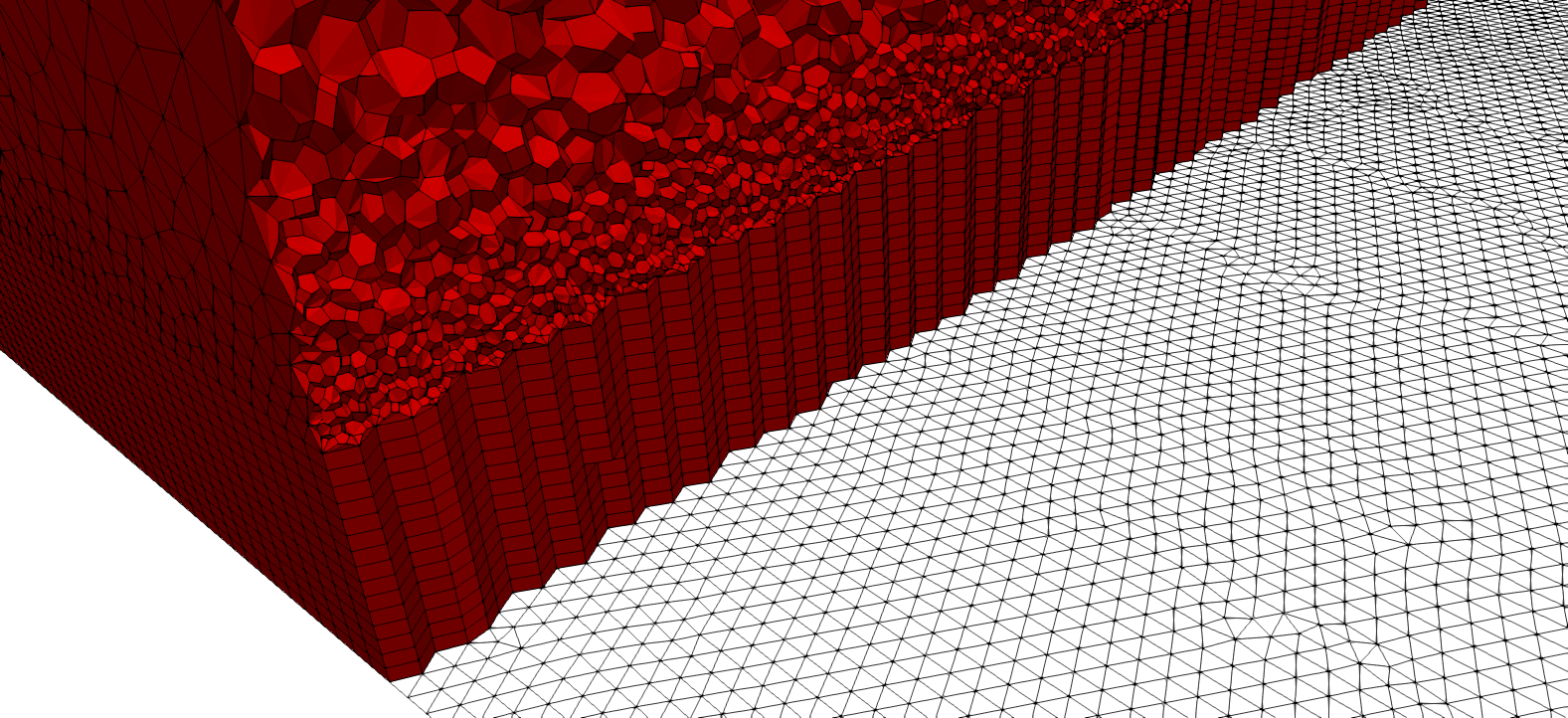}}
	\caption{Structure of the three grids used in the TBL simulations. In each plot, the white elements indicate the wall-surface grid, and the volume grid is cut for visualization.}
	\label{fig:tbl_grids}
\end{figure}

To generate the volume meshes, Star-CCM+\textsuperscript{\textregistered} was used.
The result is shown in Figure~\ref{fig:tbl_grids}, featuring a spanwise cut through each of the grids in order to expose their structure.
To avoid possible confusion, it is stressed the quad surface mesh is unstructured, although a significant part of the wall surface is covered by regular squares.
A major difficulty has been prescribing the variation of the height of the extruded prismatic layers with $x$.
Unfortunately, supplying an analytical relationship as input is not supported, and the only straight-forward option to locally control the height is using `volumetric controls'.
Each control allows to specify the height's value within the boundaries of a particular shape, for example, a box.
The approach used was then to define boxes partitioning the whole domain in the streamwise direction and prescribing the layer height in each box.
As a result, a piecewise-constant distribution of the height following the analytical estimate could be achieved.
To minimize the size of the jump in the height across neighbouring boxes, a large number of boxes, 160, was used.
An important reason for why the variation of the layer height should be as smooth as possible is that the jump in the total height leads to a jump in the height of the first few cells, and one of them (here, the second), serves as the sampling point for the wall model.
A jump in $h$, in turn, leads to a slight jump in the predicted wall shear stress.
Here, the distribution of $h$ is sufficiently smooth to avoid such kind of artefacts in the solution, see Figure~\ref{fig:tbl_cf}.

The total number of cells in each grid is provided in Table~\ref{tab:tbl_grids}.
Additionally, its decomposition into the cells covering the TBL and the freestream regions is given.
Looking at the former, the expected grid size relationship between quad-, poly- and tri-prisms is observed: the tri-prism grid contains the least amount of cells, and the poly-prism the highest.
For the poly- and tri-prism grids the freestream region is covered by $\approx 3$ million cells, but for the quad-prism grid the number is higher, $\approx 4.78$ million, which makes the total size of the quad- and poly-prism grids almost equal, with the quad-prism grid being slightly larger.

\begin{table}[htp!]
	\caption{The total number of cells, $N^\text{total}_\text{cells}$, in millions, in the grids used for the TBL simulations. The corresponding number of prismatic cells covering the TBL region, $N^\text{TBL}_\text{cells}$, and the freestream region, $N^\text{FS}_\text{cells}$. The number of hours, in thousands, it would take to advance each simulation by one $T_f$ on a single core, $N_\text{hours}$.}
	\label{tab:tbl_grids}
	\begin{center}
		\begin{tabular}{l l l l l}
			\toprule
			& $N^\text{total}_\text{cells}$ & $N^\text{TBL}_\text{cells}$ & $N^\text{FS}_\text{cells}$ & $N_\text{hours}$ \\
			\midrule
			Quads  & $11.23$ & $6.45$ & $4.78$ &  $4.3$ \\
			Polys  & $11.19$ & $8.03$ & $3.15$ &  $4.3$ \\
			Tris   & $8.59$  & $5.59$ & $3.00$ &  $3.6$ \\
			\bottomrule
		\end{tabular}
	\end{center}
\end{table}

The cells in the freestream region are heavily clustered in a thin slice directly above the prismatic layers, see Figure~\ref{fig:tbl_grids}.
In Star-CCM+\textsuperscript{\textregistered}, their total number is heavily influenced by the `volume growth rate' user input parameter, which defines the rate of transition in the cell volume from the prismatic layers towards the top boundary.
The same value of the growth rate was prescribed for all three grids, but in spite of that more cells were generated in the case of quad-prisms.
It is difficult to say whether this is an artefact of the mesh generation algorithm or if it is fundamentally more difficult to couple a generic polyhedral grid with a quad-dominated surface mesh.
The important point is that the cells constituting the transition region from the TBL to the freestream may account for a significant portion of the mesh and care should be taken in defining the properties of the grid in this region.
On the one hand, it is attractive to keep the amount of cells in the transition region to a minimum to reduce simulation time.
On the other hand, it represents a buffer that could be useful in case the values $\delta$ are underestimated.

Table~\ref{tab:tbl_grids} also reports the simulation time metric $N_\text{hours}$.
As expected, the simulation on the tri-prism grid is the cheapest.
The quad- and poly-prism grid simulations $N_\text{hours}$ is roughly the same.
Recall that a good a-priori metric of the computational complexity is the number of faces.
The prismatic layers of the quad-prism grid contain fewer faces than that of the poly-prism, but this difference is compensated by the larger number of cells in the freestream region, where the cells have a large number of faces due to their arbitrary polyhedral shapes.

\subsubsection{Results} \label{sec:tbl_results}

The results of the conducted simulations are now presented.
We note that, compared to the channel flow case, comparison with benchmark data is more difficult, because the outer scales of the flow are not defined by the geometry of the computational domain.
As a consequence, the choice of the scales and the procedure to compute them can have a significant influence on the level of agreement with reference data.
For the outer length scale, the most robust choice is generally considered to be the momentum thickness,~$\theta$.
However, in the case of WMLES this is not necessarily the case, since computing $\theta$ involves integration across the whole mean velocity profile, which can exhibit erroneous values in the inner layer.
Additionally,~$\theta$ depends on the value of the free stream velocity $U_0$, which also serves as the outer velocity scale.
In principle, $U_0$ is a fixed input parameter, however, in practice, the mean velocity profiles may locally exhibit values larger than $U_0$, and it is thus more robust to compute $U_0$ as function of $x$ as the local maximum of $\mean{u(y)}$.

To simultaneously evaluate the procedures to compute the outer scales and the growth rate of the boundary layer, the evolution of $\rey_\theta = \theta U_0/\nu$ in the streamwise direction is considered.
The following power-law estimate connecting $\rey_\theta$ and $\rey_x$ is used as reference data~\cite{Rezaeiravesh2016}

\begin{equation}
\label{eq:power_law_retheta}
	\rey_\theta =  0.0167 \cdot \rey_x^{0.846} + 373.83.
\end{equation}

\begin{figure}[htp!]
	\centering
	\includegraphics[center]{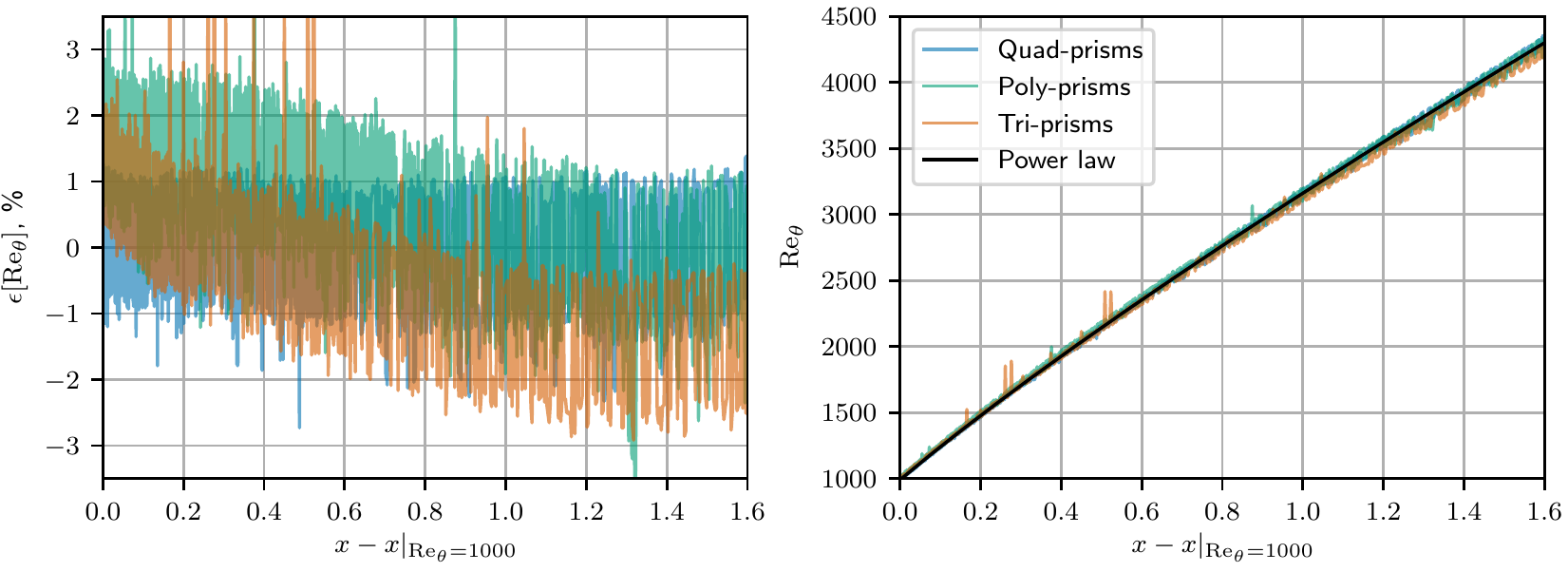}
	\caption{\textit{Left}: The relative errors in the values of $\rey_\theta$ obtained in the TBL simulations, as compared to a power-law estimate. \textit{Right}: The values of $\rey_\theta$ obtained in the TBL simulations and the reference power-law estimate.}
	\label{fig:tbl_retheta}
\end{figure}

To compare the growth across the three simulations, the origin of the $x$-axis is shifted to the point where $\rey_\theta \approx 1000$ for each of the simulations.
The computed relative errors in $\rey_\theta$ with respect to the power law are shown in Figure~\ref{fig:tbl_retheta}.
Due to the difficulties in robustly computing $\theta$ discussed above they exhibit high-frequency oscillations.
The agreement with the power law is remarkably good, the magnitudes of the errors rarely exceed 3\%.
No significant dependency on the type of employed grid is observed.
The values of $\rey_\theta$ themselves along with the power-law estimate employed are shown in the right plot of Figure~\ref{fig:tbl_retheta}.

The accurate prediction of $\rey_\theta$ warrants using it as the independent variable when analysing the distributions of other flow quantities across the length of the plate.
However, it is beneficial to first filter out the high-frequency errors in the obtained data so that they do not propagate into the subsequent results.
To that end, the raw $\rey_\theta$ values were least-square fit to a power-law function based on~\eqref{eq:power_law_retheta}: $a_0 \cdot \rey_x^{0.846} + a_1$, where $a_0$ and $a_1$ are determined by the fit.
In the remainder of this section, $\rey_\theta$ will refer to the fitted values of the quantity.

\begin{figure}[htp!]
	\centering
	\includegraphics[center]{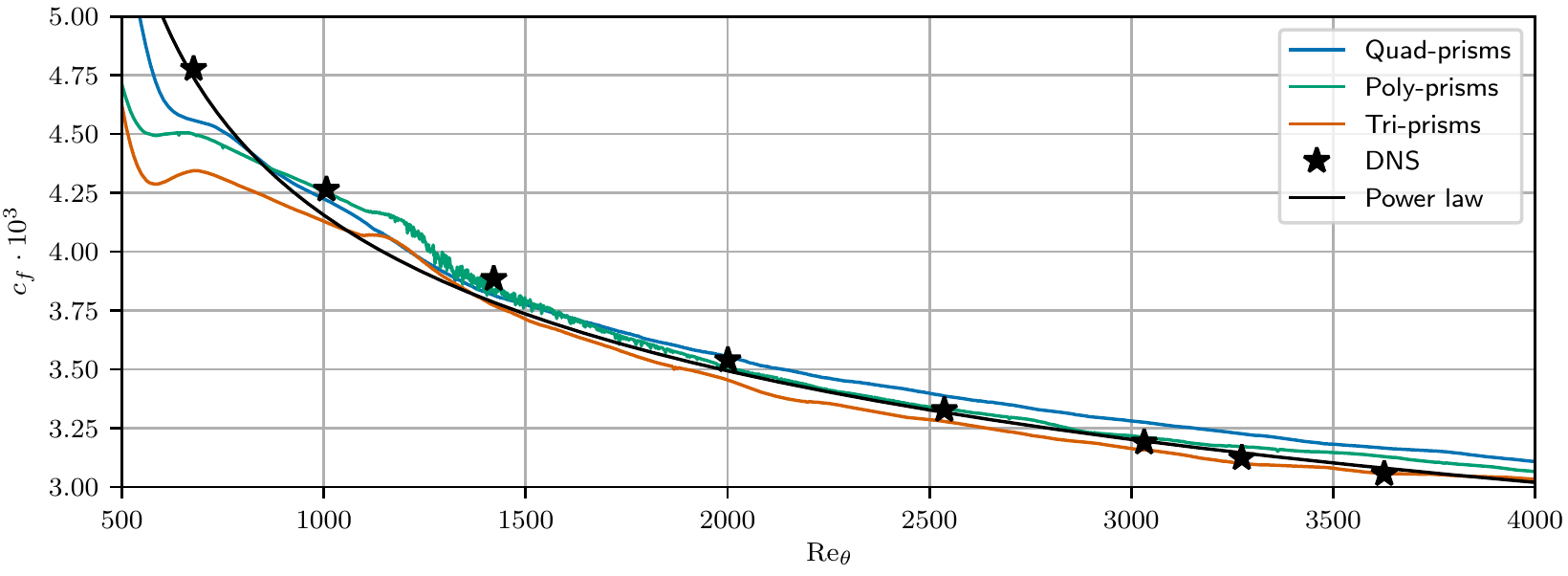}
	\caption{The values of the skin-friction coefficient, $c_f$, obtained in the TBL simulations. }
	\label{fig:tbl_cf}
\end{figure}

Attention is now turned to the performance of the wall model, which is evaluated by considering the values of the skin-friction coefficient, $c_f$.
The obtained values are shown in Figure~\ref{fig:tbl_cf}.
The DNS data by Schlatter and Örlü~\cite{Schlatter2010} and a power law estimate from~\cite{Rezaeiravesh2016} are used as reference.
Excellent agreement with both is found.
The results from all three grids exhibit a similar level of accuracy, however on the poly-prism grid some oscillations in the values of the wall-shear stress are observed up to $\rey_\theta \approx 2000$, as reflected in the corresponding $c_f$ curve in the figure.

\begin{figure}[htp!]
	\centering
	\includegraphics[center]{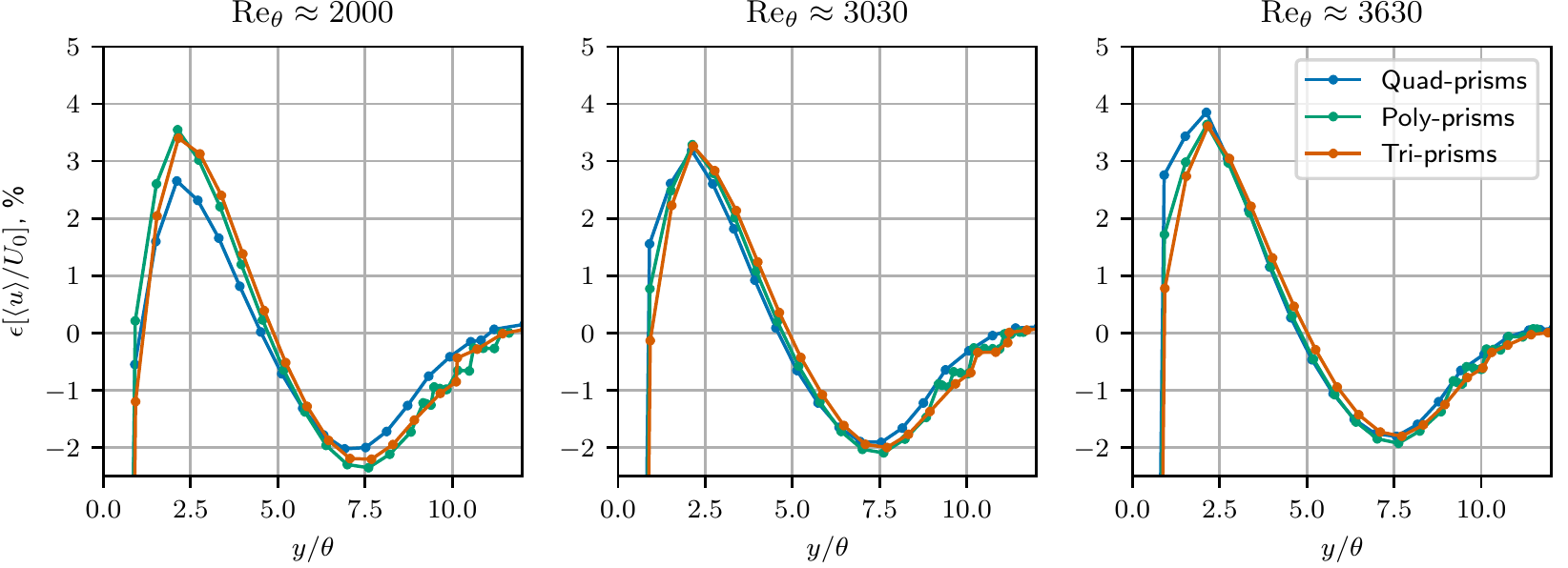}
	\caption{Error profiles in the mean streamwise velocity in TBL simulations}
	\label{fig:tbl_u}
\end{figure}

In Figure~\ref{fig:tbl_u}, the obtained relative error profiles of the mean streamwise velocity are presented at selected values of $\rey_\theta$.
DNS data~\cite{Schlatter2010} is used as reference, and similarly to the WMLES data, the velocity is normalized with its maximum value at the considered streamwise location.
Consequently, the values of the errors go towards zero at the edge of the TBL.
The curves from the three simulations are very similar.
Inspection of the values close to the free stream reveals slight oscillations in the solution.
They are caused by the fact that the boundary layer extends above the prismatic layers.
Agreement with the DNS is reasonably good and comparable to what is reported for channel flow at a similar resolution, see Figure~\ref{fig:channel_u}.
The error pattern is also similar.
The reference data is first over-predicted and towards the edge of the TBL under-prediction is observed instead.

\begin{figure}[htp!]
	\centering
	\includegraphics[center]{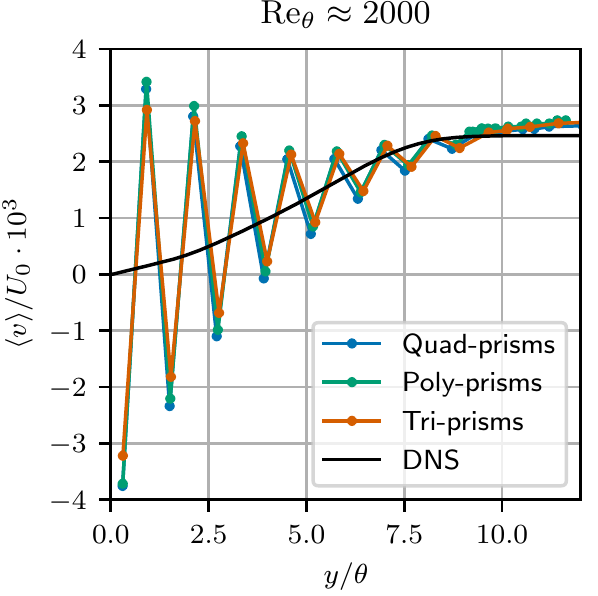}
	\caption{Profiles of the mean wall-normal velocity in TBL simulations.}
	\label{fig:tbl_v}
\end{figure}

Unlike channel flow, in the case of the TBL the mean wall-normal velocity is also non-zero, although several orders of magnitude smaller than the streamwise component.
Unfortunately, this quantity could not be predicted reliably in the conducted WMLES, and the obtained fields exhibit heavy oscillations, see Figure~\ref{fig:tbl_v}.
It is speculated that a denser grid is necessary to diminish them.
It is interesting that in spite of these oscillations the growth of the TBL could be predicted correctly.

\begin{figure}[htp!]
	\centering
	\includegraphics[center]{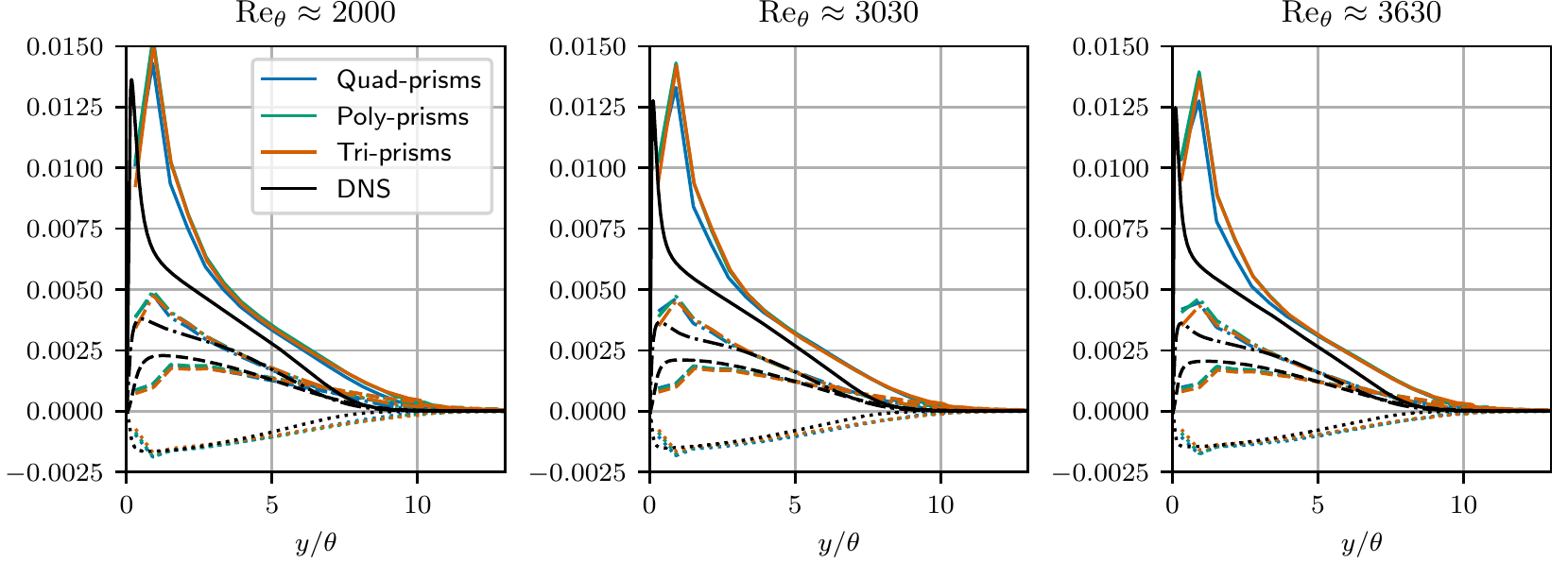}
	\caption{Profiles on non-zero components of the Reynolds stress tensor obtained in the TBL simulations. Solid lines: $\mean{u'u'}$; dashed lines: $\mean{v'v'}$; dash-dotted lines: $\mean{w'w'}$; dotted lines: $\mean{u'v'}$.}
	\label{fig:tbl_rey}
\end{figure}

The profiles of the components of the Reynolds-stress tensor are analysed next.
We do not consider the relative errors here due to the fact that the stresses decay to zero quite fast towards the edge of the TBL, and the relative error becomes a poor metric.
The obtained results are shown in Figure~\ref{fig:tbl_rey}.
As with the other quantities, the accuracy appears to be independent of the type of grid employed.
The least accurately predicted component is $\mean{u'u'}$, whereas the other two diagonal components are captured quite well.
The turbulent shear stress exhibits a different error pattern compared to channel flow.
This is due to the fact that here $\mean{u'v'}$ is not directly coupled to the values of $u_\tau$.
The near-wall peak is predicted quite accurately, but the values further above are slightly under-predicted.

\begin{figure}[htp!]
	\centering
	\includegraphics[center]{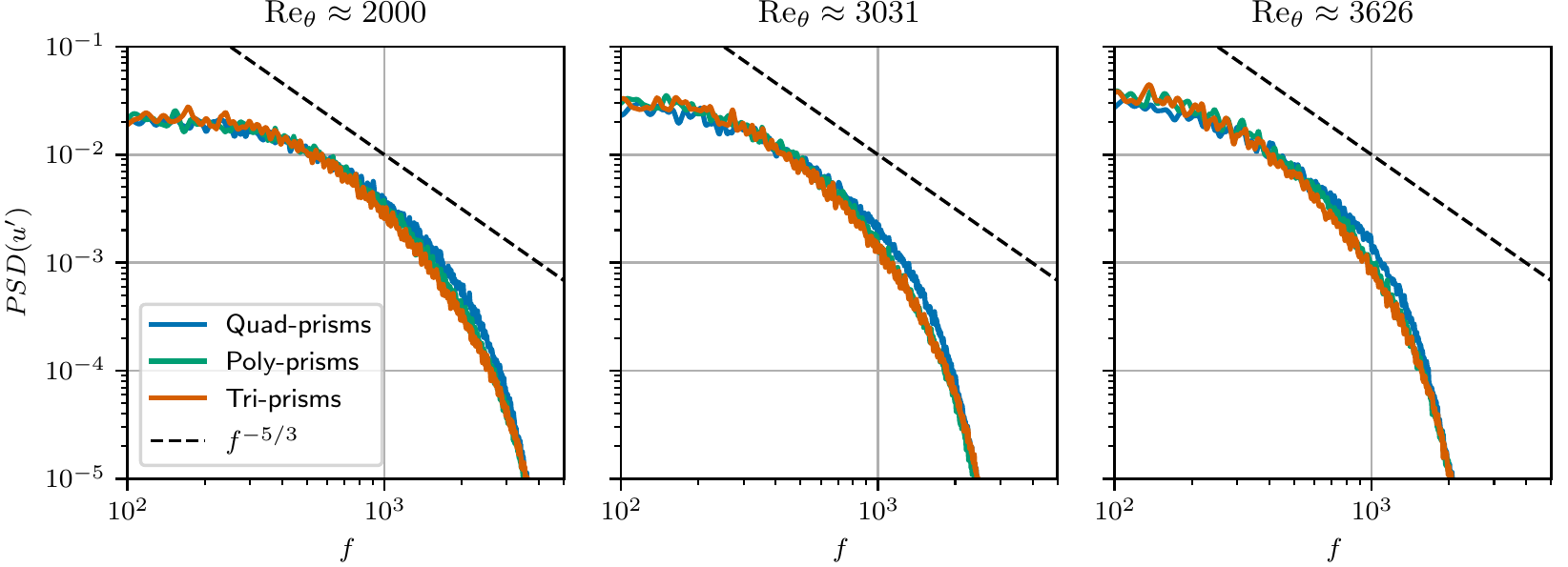}
	\caption{Power spectral densities of the streamwise velocity in the TBL simulations.}
	\label{fig:tbl_time_spectra}
\end{figure}

Finally, the PSDs of the streamwise velocity fluctuations are considered.
The same computational procedure as for channel flow has been used to obtain the estimates.
Fifty probes uniformly distributed along the spanwise direction were used to collect the data for each considered streamwise location.
The wall-normal position of the probes is $y\approx0.5\delta(x)$.
The results are shown in Figure~\ref{fig:tbl_time_spectra}, with the plots zoomed in on the inertial subrange.
No substantial dependency of the PSD on the grid type is observed, although a marginally smaller level of dissipation is obtained on the quad-prism grid.

\subsubsection{Summary} \label{sec:tbl_summary}
Largely, the outcomes of the TBL study confirm what was previously observed for channel flow.
All three unstructured grids led to very similar results with respect to all the considered quantities.
Agreement with WMLES results of the same flow reported in~\cite{Mukha2018a} is also good.
Furthermore, even though flat-plate TBL flow is more difficult to predict than channel flow due to the distribution of $\delta$ being a simulation outcome instead of a geometric parameter, the achieved accuracy is similar.
Since no grid type led to improved predictions, it can be argued that the selection should be made based on computational cost, meaning that the tri-prism grid constitutes the optimal choice.
Finally, the inability of the WMLES to capture $\mean{v}$ raises a concern.
It would be interesting to see if the issue persists at grid resolutions higher than $d=\delta/15$.
\section{Conclusions} \label{sec:conclusions}

This study presents a careful evaluation of the predictive accuracy of WMLES on unstructured grids.
To that end, we propose: \textit{i)} A mesh construction strategy that relates the resolution of the mesh to the local outer scale of the flow. \textit{ii)} A mesh resolution metric based on cell-centre distance that is suitable for comparative analysis of results obtained on meshes with different cell topology.

Results from channel flow and flat-plate TBL simulations clearly show that unstructured prismatic meshes, constructed according to the proposed strategy, can be used for WMLES without any significant impediment to accuracy as compared to results on structured hexahedral meshes.

Three candidates for the base shape of the prisms are considered in the study: triangles, quadrilaterals and hexagons.
The choice of the shape did not have a significant impact on the results, although in the case of channel flow using triangular prisms did lead to slight deterioration of accuracy.
Consequently, the selection of the shape can be guided by practical concerns, of which, arguably, the most important one is the simulation cost.
From that perspective, triangular prisms constitute the superior choice because it minimizes both the number of cells and the number of faces per cell.
Another important guideline for choosing the base shape is simply the quality of the produced grid for the geometry at hand.
Here it was only necessary to discretize planar surfaces, therefore excellent quality was easy to achieve with all the considered shape options, but for a practical case the situation is likely to be different.

A possible direction of future work is to see how strongly the accuracy of the results deteriorates when the quality of the prisms is worsened, i.e.~when the underlying surface mesh contains skewed and non-orthogonal cells.
Also, robustness with respect to the dissipativity of the employed numerics would be interesting to study.
Both to see whether good accuracy can still be achieved with more dissipative schemes, and whether simulations could be kept stable when less dissipation is introduced.
The next step in the complexity of the flow problem would be TBLs over curved surfaces and affected by pressure gradients.

%
%
%
%
%

\section{Acknowledgements}
This work was supported by grant number P38284-2 from the Swedish Energy Agency.
The simulations were performed on resources at Chalmers Centre for Computational Science and Engineering (C3SE).

\bibliographystyle{plain}
\bibliography{../../library}


\end{document}